\documentclass[sigconf]{acmart}
\usepackage{pifont}
\usepackage{amsthm}
\usepackage{subcaption}
\usepackage{algorithm}
\usepackage{algpseudocode}
\usepackage[table]{xcolor}
\usepackage{multirow}
\usepackage{balance}
\usepackage{fontawesome5} % 提供 \faIcon 命令
\theoremstyle{definition} % 使用定义样式（标题加粗，正文正体）
\newtheorem{assumption}{Assumption}
\newtheorem{remark}{Remark}
\newcommand{\eb}[1]{{\scriptsize$\pm#1$}}
\definecolor{darkgreen}{RGB}{0, 100, 0}
\AtBeginDocument{%
  }

\copyrightyear{2026}
\acmYear{2026}
\setcopyright{cc}
\setcctype{by}
\acmConference[MM '26]{Proceedings of the 34th ACM International Conference on Multimedia}{November 10--14, 2026}{Rio de Janeiro, Brazil}
\acmBooktitle{Proceedings of the 34th ACM International Conference on Multimedia (MM '26), November 10--14, 2026, Rio de Janeiro, Brazil}
\acmDOI{10.1145/3767308.3834936}
\acmISBN{979-8-4007-2213-4/2026/11}

\begin{document}

\title{M$^3$TR: Temporal Retrieval Enhanced Multi-Modal Micro-video Popularity Prediction}

\author{Jiacheng Lu}
\authornote{Both authors contributed equally to this research.}
\email{jack111@sjtu.edu.cn}
\orcid{0009-0003-2479-2620}
\author{Weijian Wang}
\authornotemark[1]
\email{SekaiWwj@sjtu.edu.cn}
\affiliation{%
  \institution{Shanghai Jiao Tong University}
  \city{Shanghai}
  \state{Shanghai}
  \country{China}
}

\author{Mingyuan Xiao}
\affiliation{%
  \institution{Shanghai Jiao Tong University}
  \city{Shanghai}
  \state{Shanghai}
  \country{China}
}
\email{xiaomy\_mark@sjtu.edu.cn}

\author{Yang Hua}
\affiliation{%
  \institution{Queen's University Belfast}
  \city{Belfast}
  \state{Northern Ireland}
  \country{UK}
}
\email{Y.Hua@qub.ac.uk}

\author{Tao Song}
\authornote{Corresponding author}
\affiliation{%
  \institution{Shanghai Jiao Tong University}
  \city{Shanghai}
  \state{Shanghai}
  \country{China}
}
\email{songt333@sjtu.edu.cn}

\author{Bo Peng}
\email{pengbo\_michael@sjtu.edu.cn}
\author{Cheng Hua}
\email{cheng.hua@sjtu.edu.cn}
\affiliation{%
  \institution{Shanghai Jiao Tong University}
  \city{Shanghai}
  \state{Shanghai}
  \country{China}
}

\author{Haibing Guan}
\affiliation{%
  \institution{Shanghai Jiao Tong University}
  \city{Shanghai}
  \state{Shanghai}
  \country{China}
}
\email{hbguan@sjtu.edu.cn}

\renewcommand{\shortauthors}{Jiacheng Lu et al.}

\begin{abstract}
Accurately predicting the popularity of micro-videos is a critical but challenging task, characterized by volatile, `rollercoaster-like' engagement dynamics. Existing methods often fail to capture these complex temporal patterns, leading to inaccurate long-term forecasts. This failure stems from two fundamental limitations: \ding{172} a superficial understanding of user feedback dynamics, which overlooks the mutually exciting and decaying nature of interactions such as likes, comments, and shares; and~\ding{173} retrieval mechanisms that rely solely on static content similarity, ignoring the crucial patterns of how a video's popularity evolves over time. To address these limitations, we propose \textbf{M$^3$TR}, a \textbf{T}emporal \textbf{R}etrieval enhanced \textbf{M}ulti-\textbf{M}odal framework that uniquely synergizes fine-grained temporal modeling with a novel temporal-aware retrieval process for \textbf{M}icro-video popularity prediction. At its core, M$^3$TR introduces a Mamba-Hawkes Process (MHP) module to explicitly model user feedback as a sequence of self-exciting events, capturing the intricate, long-range dependencies within user interactions (for \textbf{limitation} \ding{172}). This rich temporal representation then powers a temporal-aware retrieval engine that identifies historically relevant videos based on a combined similarity of both their multi-modal content (visual, audio, text) and their popularity trajectories (for \textbf{limitation} \ding{173}). By augmenting the target video's features with this retrieved knowledge, M$^3$TR achieves a comprehensive understanding of prediction. Extensive experiments on two real-world datasets demonstrate the superiority of our framework. M$^3$TR achieves state-of-the-art performance, outperforming previous methods by up to \textbf{19.3}\% in nMSE and showing significant gains in addressing long-term prediction challenges.
\end{abstract}




\received{20 February 2007}
\received[revised]{12 March 2009}
\received[accepted]{5 June 2009}

\maketitle

\section{Introduction}
\label{sec:intro}

Micro Video Popularity Prediction (MVPP) has become a critical area of research and application with the rapid rise of short-form video platforms such as TikTok, YouTube Shorts and Instagram Reels~\cite{MVPP}. MVPP aims to forecast the popularity of micro-videos based on content, context and user interactions. It is essential for enhancing recommendation systems~\cite{recommendation1,recommendation2}, enabling public sentiment analysis and control~\cite{governance}, optimizing targeted advertising~\cite{ad}, and guiding content creation strategies. MVPP offers multifaceted advantages, from empowering users to navigate the vast digital landscape more effectively to enhance applications like personalized marketing, social trend analysis, and content moderation~\cite{MVPP}. In the context of this work, MVPP is defined not only as forecasting a single, static score, but as predicting the \textit{temporal evolution} of a video's user engagement metrics: likes, shares, and comments.

\begin{figure}
    \centering
    \includegraphics[width=\columnwidth]{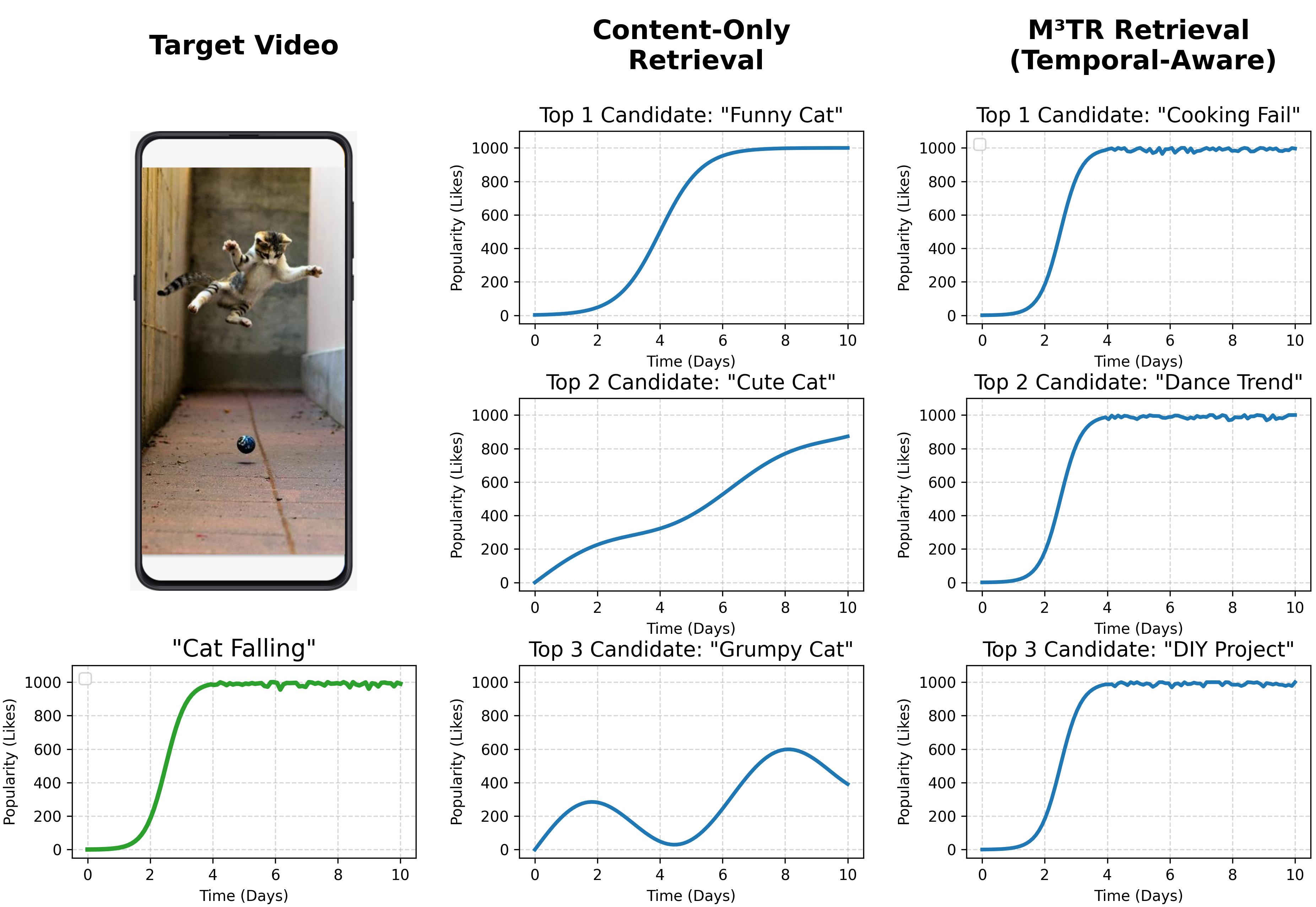}
    \caption{Illustration of the core motivation for M$^3$TR. 
        Given a target video (left) exhibiting a ``flash-in-the-pan" popularity trajectory, a conventional \textbf{Content-Only Retrieval} system (middle) identifies videos that are thematically similar (e.g., other cat videos) but whose popularity patterns are fundamentally different, offering poor predictive value. 
        In stark contrast, our \textbf{M$^3$TR Retrieval} (right), which is temporal-aware, retrieves videos with entirely different content (e.g., ``cooking fail" and ``dance trend") but nearly identical popularity trajectories. This demonstrates that the temporal user feedback of videos is a critical predictive signal that is missed by content-centric approaches.}
    \Description{A three-part comparison of retrieval strategies. On the left, a target video showing a cat is paired with a popularity trajectory that rises rapidly and then reaches a plateau. The center column shows three videos retrieved using content similarity; all are cat-related, but their popularity curves have substantially different shapes, including gradual growth and non-monotonic variation. The right column shows three videos retrieved by M3TR, including videos with unrelated visual content such as cooking, dancing, and a do-it-yourself project. Despite their different content, their popularity curves exhibit temporal patterns similar to that of the target video, with rapid growth followed by saturation.}
    \label{fig:motivation}
\end{figure}

Previous research in MVPP can be broadly categorized into three main approaches: feature-based modeling, historical sequence modeling, and similarity-based modeling. Early feature-based methods focus on manually extracting static cues (e.g., image brightness and audio volume) but struggle with scalability and the complexity of modern content. More recent deep learning approaches have advanced behavior modeling by integrating user engagement metrics (e.g., likes and shares) with multi-modal features. However, these methods often treat user feedback as a simple aggregated time series,  which leads to the \textbf{limitation}~\ding{172}. They largely overlook the fine-grained, mutually exciting nature of user interactions, where a burst of likes might trigger a subsequent wave of shares, or negative comments might inhibit future growth. 

To overcome the problems of learning from a single video's data, retrieval-based methods have gained attention. These approaches enhance predictions by leveraging a vast repository of historical videos, retrieving similar content to inform the forecast~\cite{MMRA,RAH}. While promising, the effectiveness of these methods is often capped by a reliance on static, content-centric similarity. As shown in Figure~\ref{fig:motivation}, a typical retrieval system might correctly match a target video of a cat with other cat videos. However, it fails to recognize that their differing popularity dynamics offer poor, or even misleading, predictive signals. The system completely misses thematically different videos (e.g., a ``dance trend" or ``DIY project") that share an identical viral pattern---a crucial piece of information for an accurate forecast. The \textbf{limitation}~\ding{173} stems from their inability to measure and retrieve based on temporal similarity, leaving a significant source of predictive power untapped (Appendix \fcolorbox{red}{white}{A}). 
%(more examples shown in Appendix \fcolorbox{red}{white}{A}).

To address these two fundamental limitations mentioned above, we propose \textbf{M$^3$TR}, a novel \textbf{T}emporal \textbf{R}etrieval enhanced \textbf{M}ulti-\textbf{M}odal framework for \textbf{M}icro-video popularity prediction. M$^3$TR introduces a paradigm shift by positing that accurate prediction requires synergizing deep temporal dynamic modeling with a temporal-aware retrieval process. Our contributions are threefold:

\textit{(i)} We propose a novel framework, M$^3$TR, that for the first time, integrates a sophisticated temporal event model directly into a multi-modal retrieval pipeline, bridging the gap between user behavior modeling and retrieval augmentation. \textit{(ii)} We introduce a (1) \textbf{Mamba-Hawkes Process (MHP) module} to explicitly capture the long-range, self-exciting, and mutually-influential dynamics within user feedback sequences, providing a far richer temporal representation than previous methods. This representation is then leveraged by our (2) \textbf{temporal-aware retrieval engine}, which computes similarity based on both multi-modal content and the temporal patterns captured by the MHP. This integrated approach uniquely enables the discovery of videos that share similar popularity trajectories, regardless of their underlying content. \textit{(iii)} We conduct extensive experiments on two large-scale, real-world datasets, demonstrating that M$^3$TR significantly outperforms state-of-the-art methods by up to \textbf{19.3}\%.

\section{Related Work}
\label{Sec:related works}

Current research primarily focuses on three approaches: feature-based modeling, historical sequence modeling, and similarity-based modeling. 

Feature extraction-based methods aim to derive information from static and dynamic attributes of videos, such as titles, descriptions, user comments and likes~\cite{VSCNN}. Earlier approaches used simple models like CNNs or RNNs for low-level features such as brightness, contrast, and motion. Despite that feature fusion and optimization like XGBoost could enhance prediction~\cite{XGB}, these global features are often proved insufficiently explanatory. After that, subsequent studies employed more sophisticated models such as Bi-LSTM~\cite{Bi-LSTM-GRU} to extract higher-order features like emotional content, object interactions, and facial expression representations. Although effective in specific contexts, these high-level features face challenges with generalizability and conditional dependencies. Recently, multi-modal integration and end-to-end modeling have emerged as a major trend. By employing attention mechanisms, these models align and fuse data from different modalities to capture complex contextual information~\cite{UGCMR, CPRP-CNN}. For instance, the AMPS ~\cite{AMPS} integrate features across multiple time windows, effectively combining global and local data to improve prediction accuracy.

Historical sequence  modeling leverages the analysis of past viewing data and user behavior to capture temporal trends in video popularity. Traditional time series methods, such as ARIMA~\cite{arima}, are commonly employed for short-term forecasting but struggle with complex nonlinear patterns. To address this, researchers have proposed contextual popularity modeling approaches that extract inherent patterns from viewing data to predict future trends~\cite{Non-Param}. Additionally, social network-based approaches utilize graph analysis to model the dynamic dissemination of videos within networks, forecasting their future popularity trajectories~\cite{Social-Forecast}. However, these methods only focus on social network analysis and lack the integration with intrinsic video content features which in fact limits their applicability.

Retrieval-based modeling improves popularity prediction accuracy by analyzing video relationships, which have also gained prominence in recent years. For example, the MMRA method combines visual and textual information to retrieve similar videos from the memory network~\cite{MMRA}. RAGTrans~\cite{RAH} employs a multi-modal memory database and hypergraph-based enhancements to generate robust representations of user-generated content, thereby improving social media popularity forecasts. These methods inherently offer interpretability by grouping similar content clusters, but they fail to exploit temporal patterns of videos with similar user feedback.

%%%%%%%%%%%%%%%%%%%%%%%%%%%%%%%%%%%%%%%
\begin{figure*}[htbp]
    \centering
    \includegraphics[width=2\columnwidth]{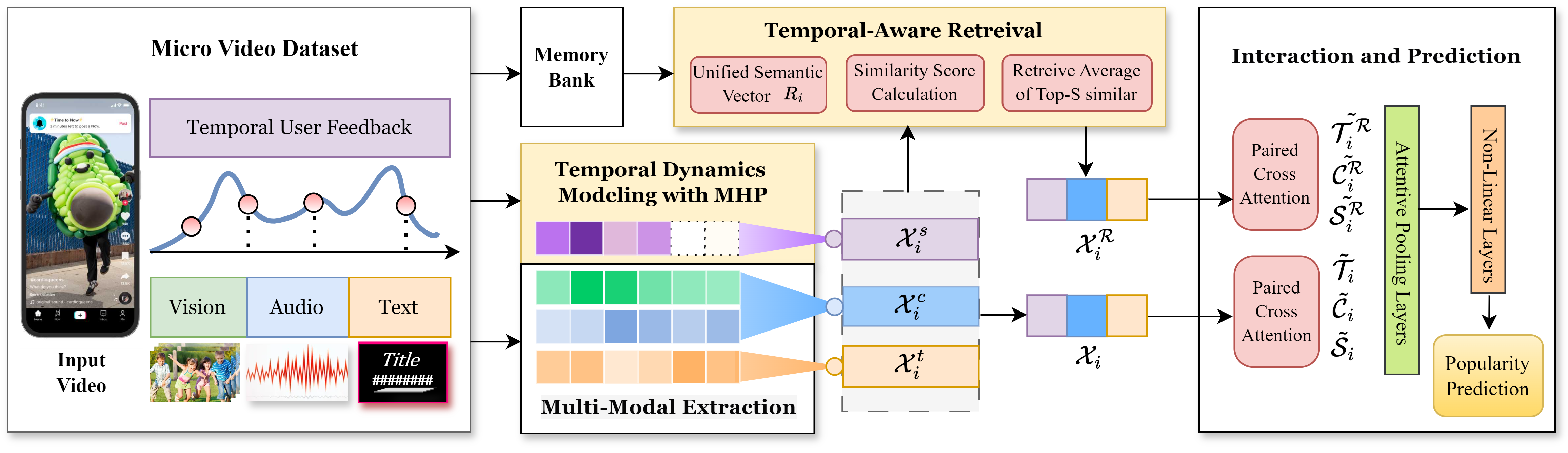}
    \caption{Overall framework. The workflow begins with two input streams: the Input Video, from which multi-modal features (vision, audio, text) are extracted, and its corresponding temporal user feedback. M$^3$TR employs a Multi-Modal Extraction module for vision, audio and text. Meanwhile, it applies \textbf{Temporal Dynamics Modeling with a Mamba-Hawkes Process (MHP)} for temporal user feedback, which explicitly captures the long-range, self-exciting nature of user interactions, generating a sophisticated temporal feature representation ($X_i^s$). This rich temporal feature powers the other core innovation of our framework: \textbf{the Temporal-Aware Retrieval engine}. Unlike conventional methods that rely on static content similarity, our engine queries the Memory Bank to identify historical videos based on a combined similarity of both their popularity trajectories (derived from $X_i^s$) and their multi-modal content. It retrieves and averages the features of the most relevant historical examples to form an augmented feature vector ($X_i^R$). Finally, in the Interaction and Prediction stage, the model fuses the video's original multi-modal features ($X_i$) with the retrieved temporal-aware features ($X_i^R$) using Paired Cross Attention. The resulting representation is then refined through Attentive Pooling and Non-Linear Layers to generate the final Popularity Prediction.}
    \Description{A left-to-right architecture diagram of M3TR. The input video on the far left is decomposed into two branches. One branch contains temporal user feedback represented as a sequence of interaction events, while the other contains vision, audio, and text. The temporal branch is processed by the Mamba-Hawkes Process, and the three content modalities are processed by the multi-modal extraction module. The resulting temporal, audio-visual, and textual feature representations are arranged as three parallel feature streams. A memory bank feeds the temporal-aware retrieval module above these streams, where semantic and temporal similarities are computed and the top similar historical videos are aggregated. The retrieved representation and the original video representation then enter the interaction and prediction block on the right. Paired cross-attention modules align the representations, followed by attentive pooling and non-linear layers that produce the final popularity prediction.}
    \label{fig2}
    
\end{figure*}

\section{Methodology} 
\subsection{M$^3$TR Framework Overview}
\label{sec:framework_overview}

% \subsubsection{Problem Formulation}
% \label{sec:problem_formulation}
Let $\mathcal{V} = \{V_1, V_2, \dots, V_N\}$ be a collection of micro-videos. Each video $V_i$ is a composite of multi-modal data streams, including visual frames, audio tracks, and textual metadata (e.g., titles and descriptions). Accompanying each video is a time-stamped sequence of user interactions, such as likes, shares, and comments. The goal of Micro-Video Popularity Prediction (MVPP) is to learn a function $\psi(\cdot)$ that takes the multi-modal content and user interaction history of a video $V_i$ as input and predicts a vector of its future popularity metrics, $\hat{Y}_i = [\hat{y}_{\text{likes}}, \hat{y}_{\text{shares}}, \dots]$. 

% \subsubsection{Framework Architecture}
% \label{sec:architecture}

To tackle this challenge, we propose the M$^3$TR framework, whose overall architecture is illustrated in Figure~\ref{fig2}. M$^3$TR firstly extracts features of multi-modalities, and specifically models temporal dynamics for user feedback. Then the data flows into a temporal-aware retrieval module to enhance feature exploitation. The final step is to integrate the retrieved information with the target video’s features to make a dynamic prediction.

\subsection{Temporal Dynamics Modeling with MHP} \label{sec:mhp}
As mentioned in Sec~\ref{Sec:related works}, a core limitation of prior work is its failure to capture the rich dynamics of user engagement. However, user interactions are not independent; for example, they form a cascade of influence where likes can inspire comments, and shares can trigger new views. To model these user interactions, we treat the users' feedback sequence as a \textit{multivariate self-exciting point process}. We thereby propose the \textbf{Mamba-Hawkes Process (MHP)}, a novel architecture that captures these intricate temporal dependencies to generate a history-aware representation of a video's popularity trajectory.

\subsubsection{The Mamba-Hawkes Process Model}

Let a user's response sequence be $\mathcal{R} = \{ (t_0, \gamma_0), (t_1, \gamma_1), \dots, (t_n, \gamma_n) \}$, where $t_i$ is the timestamp and $\gamma_i$ is a one-hot vector indicating the event type (e.g., like and share). The core of our MHP model is its intensity function $\lambda_k(t)$, which defines the instantaneous probability of an event of type $k$ occurring at time $t$. We formulate this intensity to include a classical self-exciting component augmented by a dynamic, non-linear context learned by a Mamba network (refer to Appendix \fcolorbox{red}{white}{B} and Appendix \fcolorbox{red}{white}{C} for explanation):

\begin{equation}\label{eq:mhp_intensity}
\begin{split}
\lambda_k(t) &= \text{softplus}\bigr( \mu_k + \sum_{m=1}^{M} \sum_{t_j^{m} < t} \alpha_{k,m} e^{-\delta_{k,m}(t - t_j^{m})} + f_k\bigl(\mathbf{h}(t)\bigr)\bigr),
\end{split}
\end{equation}
where $\mu_k$ is the baseline intensity for event type $k$; The summation term models the linear, exponentially decaying influence of past events $t_j^{m}$ of type $m$ on the current intensity of type $k$, governed by excitation coefficients $\alpha_{k,m}$ and decay rates $\delta_{k,m}$; $f_k(\mathbf{h}(t))$ is a non-linear function whose input, $\mathbf{h}(t)$, is a history-aware hidden state produced by a Mamba network. Moreover, to strictly enforce the non-negativity constraint required by point process definitions, we apply a link function, $\lambda_k(t)=\text{softplus}(\cdot)$, to the entire intensity function. It is worth noting that the term $f_k(\mathbf{h}(t))$ allows the model to capture complex, long-range dependencies that the classical component cannot (refer to Appendix \fcolorbox{red}{white}{C.9}).

\begin{figure}[t]
    \centering
    \includegraphics[width=0.9\columnwidth]{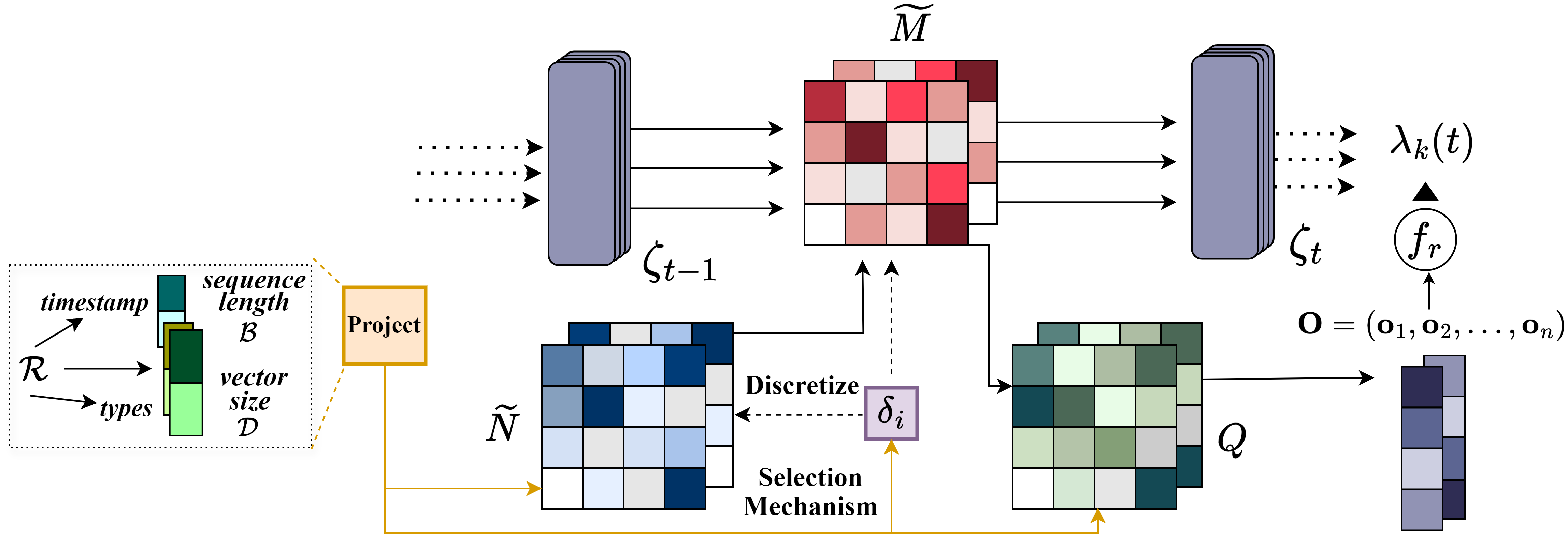} % Your figure file
    \caption{The Mamba-Hawkes Process (MHP) Architecture. It computes a dynamic intensity function by combining a classical Hawkes process with a non-linear context vector generated by a Mamba network that processes the entire event history.}
    \Description{A computational diagram of the Mamba component inside the Mamba-Hawkes Process. A time-stamped interaction event is first embedded and projected into the state-space model. The projected sequence follows two connected paths: one maintains and updates a compressed hidden state across the event history, while the other uses an input-dependent discretization and selection mechanism to determine the state-space parameters. These paths interact repeatedly as the sequence is processed. The resulting history-aware state is transformed into a non-linear context term. This context term is then combined with the classical Hawkes contribution from previous events to determine the event-type intensity at the current time.}
    \label{fig:mhp_arch}
\end{figure}

\paragraph{Dynamic Context with Mamba}
The Mamba architecture~\cite{mamba} is uniquely suited to compute the dynamic context $\mathbf{h}(t)$ due to its efficiency in modeling long sequences. As illustrated in Figure~\ref{fig:mhp_arch}, it processes the sequence of embedded event vectors, $v_{t_i} = \gamma_i E^\top$, where $E \in \mathbb{R}^{d \times M}$ is a learnable embedding matrix. Mamba maintains a compressed hidden state $\zeta_{t_i}$, which is updated at each event time $t_i$ based on the time difference $\Delta t_i = t_i - t_{i-1}$ and the current event embedding $v_{t_i}$. This update mechanism, based on a structured state-space model (SSM), allows it to selectively recall or forget information over long horizons, yielding a rich final hidden representation $\mathbf{h}(t)$ that informs the intensity function (refer to Appendix \fcolorbox{red}{white}{M} for a demo case of MHP's function). 

\subsubsection{Sequence Generation via NHPP Simulation}

Deep learning models typically require fixed-length inputs, but user interaction sequences are variably-lengthed. To address this, we treat the event sequence as a Non-Homogeneous Poisson Process (NHPP) governed by our learned intensity $\lambda_k(t)$ from Eq.~\eqref{eq:mhp_intensity}. We then employ \textbf{Ogata's thinning algorithm}~\cite{thinning} to simulate events from this process, augmenting shorter sequences to a desired fixed length in a statistically principled manner (refer to Appendix \fcolorbox{red}{white}{D}).

The thinning algorithm first establishes a constant upper bound $\lambda_{\max}^{(k)}$ for the intensity function over the simulation interval. It then generates candidate events from a simpler, computationally efficient homogeneous Poisson process with rate $\lambda_{\max}^{(k)}$. Finally, each candidate event at time $t_{\text{cand}}$ is stochastically ``thinned" (i.e., accepted or rejected) with an acceptance probability of $\lambda_k(t_{\text{cand}}) / \lambda_{\max}^{(k)}$. This procedure generates a padded event sequence whose statistical properties are faithful to our complex MHP intensity function. The final output of this module is a fixed-length temporal embedding, $\mathbf{X}_i^t$, derived from this complete event sequence, which serves as the temporal signature of video $V_i$ for our downstream retrieval task.

\subsubsection{Theory Analysis}
\label{sec:theorem1}
To ensure the statistical soundness of our temporal modeling, we formally state and prove that the estimator for the parameters of the Mamba-Hawkes Process (MHP) possesses desirable oracle properties~\cite{leeb2008sparse}. This is crucial as the MHP module forms the core of our temporal feature extraction pipeline. The ``oracle property" implies that, with enough data, our parameter estimator behaves as if it knew in advance.
\begin{theorem}[Oracle Properties of the MHP Estimator]
Let the d-dimensional counting process $N(t)$ be generated by a multivariate Hawkes process with its true conditional intensity vector governed by a sparse parameter vector $\theta^* \in \mathbb{R}^p$. Let $A = \{j | \theta_j^* \neq 0\}$ be the true active set of size $|A| = q \ll p$. The estimator $\hat{\theta}_N$ is obtained by maximizing the penalized log-likelihood from $N$ i.i.d. observed video interaction sequences over a time horizon $T$:
\begin{equation}
    \hat{\theta}_N = \arg\max_{\theta \in \mathbb{R}^p} \left\{ L_N(\theta) - \sum_{j=1}^p p_{\gamma_N}(|\theta_j|) \right\},
\end{equation}
where $L_N(\theta)$ is the normalized log-likelihood, and $p_{\gamma_N}(\cdot)$ is a non-concave penalty function such as MCP \citep{zhang2010}, with regularization parameter $\gamma_N$.

Under standard regularity assumptions (formally stated in Appendix G), the estimator $\hat{\theta}_N$ possesses the oracle property. That is, as $N \to \infty$:
\begin{enumerate}
    \item \textbf{Sparsity Recovery:} The estimator correctly identifies the true zero and non-zero parameters with probability approaching one. Formally, $P(\{\hat{\theta}_{N,j} \neq 0\} = A) \to 1$.
    \item \textbf{Asymptotic Normality:} The estimator for the non-zero parameters, $\hat{\theta}_{N,A}$, is asymptotically normal and has the same asymptotic distribution as the oracle estimator, which knows the true active set A beforehand.
    \[
    \sqrt{N}(\hat{\theta}_{N,A} - \theta_A^*) \xrightarrow{d} \mathcal{N}(0, I_{A,A}(\theta^*)^{-1}),
    \]
    where $I_{A,A}(\theta^*)$ is the Fisher information matrix for the non-zero parameters, evaluated at the true parameter $\theta^*$.
\end{enumerate}
\end{theorem}

The proof is provided in Appendix \fcolorbox{red}{white}{G}.
\subsection{Temporal-Aware Retrieval} \label{sec:retrieval}
Having established a method to generate powerful temporal embeddings in Section~\ref{sec:mhp}, we now describe how to leverage this information for prediction. A video's popularity is often correlated with that of existing, similar videos. However, ``similarity" is twofold: videos can be similar in their \textit{content} (e.g., visual themes, topics) or in their \textit{popularity dynamics} (e.g., a slow burn vs. a viral explosion). Our framework's core hypothesis is that the most relevant videos for retrieval are those that are similar on \textbf{both} axes. To this end, we design a temporal-aware retrieval module. \par
For each video $V_i$ in our dataset, we first extract a set of feature embeddings and are stored in a memory bank, representing its multi-modal content and its temporal signature.  Aligning all modalities, we generate unified semantic vectors, calculate the similarity scores and retrieves the top-$S$ most similar videos from a pre-built memory bank.

\paragraph{Multi-modal Content Embeddings ($X_i^c, X_i^{t}$).}
We employ standard, pre-trained models to encode the primary content modalities. Visual features from keyframes are processed by a Vision Transformer (ViT)~\cite{VIT}, while corresponding audio segments are processed by an Audio Spectrogram Transformer (AST)~\cite{gong2021ast}. The resulting embeddings are concatenated and passed through a linear layer to produce a unified audio-visual embedding, $X_i^c \in \mathbb{R}^{K \times d}$. Textual metadata is encoded using the AnglE sentence embedding model~\cite{aoe} to produce a text embedding, $X_i^t \in \mathbb{R}^{n_w \times d}$. Further implementation details are deferred to Appendix \fcolorbox{red}{white}{E}.%~\ref{app:extraction}.

\paragraph{Temporal Dynamics Embedding ($X_i^s$).}
The temporal signature of the video, $X_i^s$, is the fixed-length event sequence embedding generated by our Mamba-Hawkes Process module, as detailed in Section~\ref{sec:mhp}. This vector is a rich summary of the video's engagement trajectory.

\paragraph{Unified Semantic Vector ($R_i$).}
To facilitate a robust content-based similarity search, we unify the different modalities into a common textual semantic space. We use pre-trained vision-to-text BLIP-2~\cite{BLIP2} and audio-to-text CLAP~\cite{CLAP} models to generate descriptive captions for the video's visual and audio content, respectively. These generated captions are concatenated with the original textual description and encoded using AnglE to produce a single, holistic semantic retrieval vector $R_i$. This vector serves as the primary key for content-based similarity.

\paragraph{Similarity Score Calculation.}
With these representations, we compute a unified similarity score $SS(V_i, V_j)$ between a target video $V_i$ and a candidate video $V_j$ from the memory bank. This score is a composition of two distinct similarity measures. $SS_{\text{modal}}(V_i, V_j) = R_i \cdot R_j^\top$ represents the content-level similarity using the unified semantic vectors, and $SS_{\text{temp}}(V_i, V_j) = \text{mean\_pool}(X_i^s) \cdot \text{mean\_pool}(X_j^s)^\top$ 
measures the similarity in popularity dynamics using the MHP-derived temporal embeddings. \par
To balance the influence of both modalities and ensure their scales are comparable, we first apply Min-Max normalization to map each similarity score to the range $[0, 1]$. We then combine them using the L2-norm to produce the final temporal-aware similarity score:
\begin{equation}
\label{eq:combined_similarity}
SS(V_i, V_j) = \sqrt{\text{Norm}(SS_{\text{modal}})^2 + \text{Norm}(SS_{\text{temp}})^2}.
\end{equation}

\paragraph{Memory Bank Retrieval.}
Our memory bank $\mathcal{B}$ stores the complete set of feature embeddings for all videos in the training corpus:
\begin{equation}
\mathcal{B} = \{ (X_1^c, X_1^t, X_1^s, R_1, Y_1), \dots, (X_N^c, X_N^t, X_N^s, R_N, Y_N) \}.
\end{equation}
For a given target video $V_i$, we compute its similarity $SS(V_i, V_j)$ with every video $V_j \in \mathcal{B}$. We then identify the set $\mathcal{V}_i^R$ containing the top-$S$ videos with the highest similarity scores. The feature embeddings and ground-truth popularity labels of this retrieved set $\{ (X_j^c, X_j^t, X_j^s, Y_j) \}_{V_j \in \mathcal{V}_i^R}$ are then passed to our final integration module (Section~\ref{sec:integration}) to serve as augmented knowledge for the prediction task. The overall retrieval process is depicted in Figure~\ref{fig5}.

\begin{figure}[tbp]
\centering
\includegraphics[width=\columnwidth]{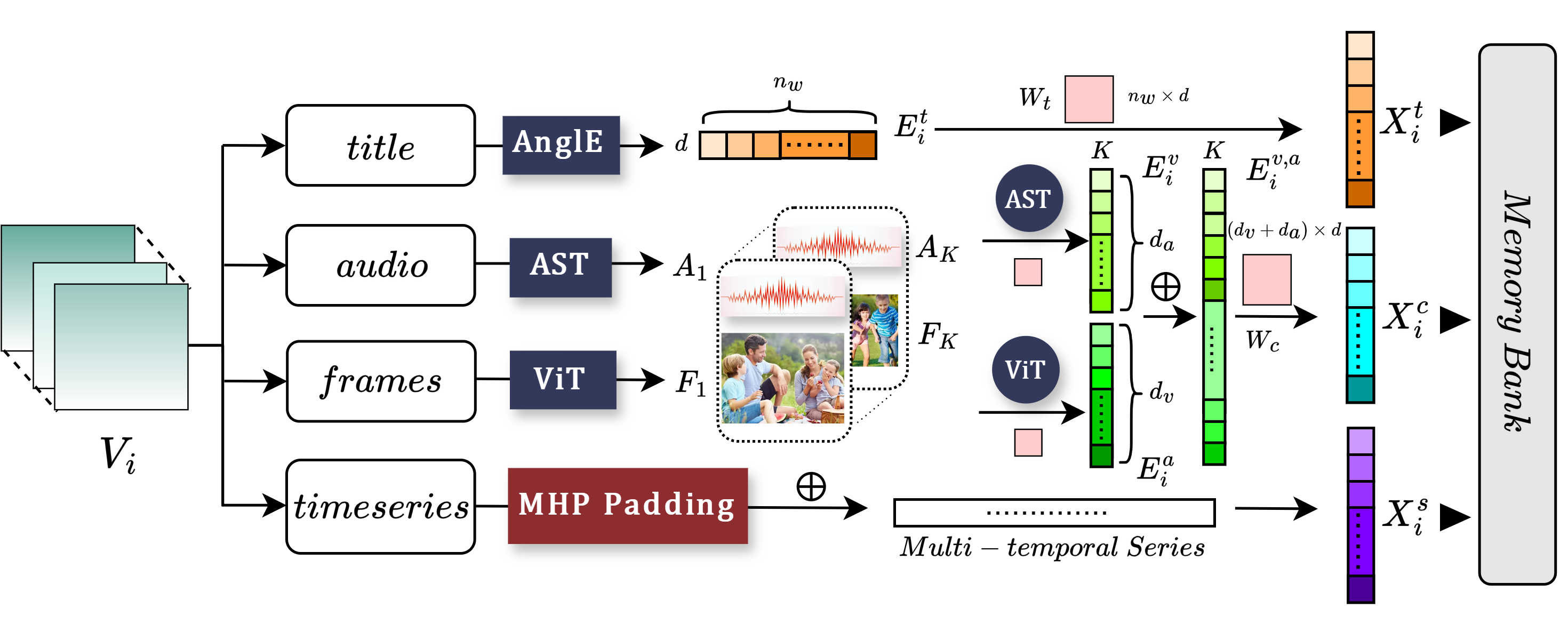}
\caption{Multi-modal/temporal Feature Extraction for Memory Bank Prebuild. Visual, acoustic, textual and time-serial features of videos are extracted and encoded into embeddings through various mechanisms. These processed embeddings are also restored for subsequent retrieval.}
\Description{A feature-extraction pipeline for constructing the memory bank. A micro-video is separated into four input streams: title, audio, video frames, and interaction time series. The title is encoded with AnglE, the audio with an Audio Spectrogram Transformer, and the video frames with a Vision Transformer. The visual and acoustic representations are
combined to form an audio-visual content representation. In parallel, the interaction time series is processed by the Mamba-Hawkes module to form a temporal representation. The resulting textual, audio-visual, and temporal feature vectors are shown as three aligned feature blocks and are stored together in the memory bank for later retrieval.}
\label{fig3}
\end{figure}

\begin{figure}
\includegraphics[width=\columnwidth]{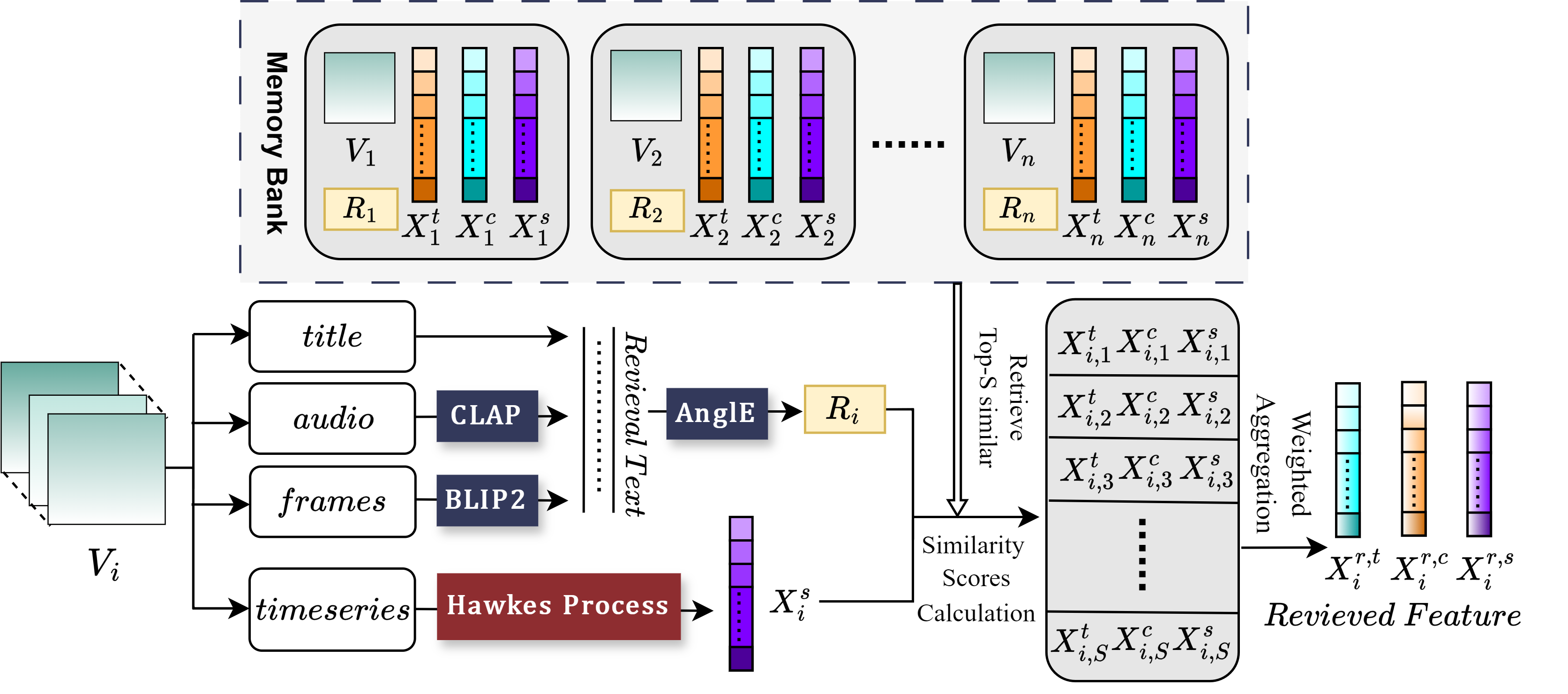}
\caption{Temporal-Aware Retrieval. Similarity scores could be computed by aligning visual, acoustic and textual features and incorporating time-serial embeddings. Analogous videos with the highest similarity scores are then retrieved from memory bank for follow-up integration process.}
\Description{A retrieval pipeline connecting a target video to a pre-built memory bank. The target video's title, audio, and frames are converted into textual semantic information using AnglE, CLAP, and BLIP-2, producing a unified semantic retrieval vector. Its interaction time series is processed separately to produce a temporal representation. For each candidate video in the memory bank, the target is compared along both semantic-content and temporal-pattern dimensions. The two similarity signals are combined into a single retrieval score. Candidates are ranked by this score, and the features of the top-ranked videos are collected and aggregated into a retrieved feature representation that is passed to the subsequent prediction stage.}
\label{fig5}
\end{figure}
 
\subsection{Interaction and Prediction} \label{sec:integration}
Once the top-$S$ similar videos are retrieved from the memory bank, the final step is to fuse this augmented information with the target video's features to make a prediction. This process involves two levels of interaction: aligning features within each video (intra-video) and then integrating information across the target and retrieved videos (inter-video).

\paragraph{Intra-Video Modal Alignment.}
To create a holistic representation for the target video $V_i$, we must first resolve potential inconsistencies between its audio-visual ($X_i^c$), textual ($X_i^t$), and temporal ($X_i^s$) embeddings. We employ a series of cross-attention networks to achieve this alignment. For instance, to enrich the textual features with visual context, we query the visual embeddings with the textual embeddings, and vice versa. This process yields a set of enhanced, modality-aligned representations ($\widetilde{C}_i$, $\widetilde{T}_i$, $\widetilde{S}_i$), where each vector is now infused with context from the other modalities. The same alignment process is applied independently to each of the $S$ retrieved videos. The detailed mechanism is provided in Appendix \fcolorbox{red}{white}{F.1}. %~\ref{app:interaction_and_prediction}.

\paragraph{Retrieval-Enhanced Fusion and Prediction.}
The core of our retrieval augmentation lies in effectively integrating the knowledge from the retrieved set, $\mathcal{V}_i^R$. We achieve this by constructing a \textit{cross-sample interaction vector} $\mathcal{I}_i$, which captures the relationships between the target video and its retrieved neighbors. This vector is formed by computing the inner product between the aligned representations of the target video and those of the retrieved videos.

Finally, all available information is concatenated into a single feature vector $H_i$:
\begin{equation}
\label{eq:H_i}
    H_i = \text{concat}\left([\widetilde{C}_i, \widetilde{T}_i, \widetilde{S}_i] \oplus \{\widetilde{C}_j^R, \widetilde{T}_j^R, \widetilde{S}_j^R, Y_j\}_{V_j \in \mathcal{V}_i^R} \oplus \mathcal{I}_i\right),
\end{equation}
where we include the aligned features of the retrieved videos as well as their ground-truth popularity labels $Y_j$ as explicit guidance. This comprehensive vector $H_i$ is fed into a non-linear layer to predict the popularity score $\hat{Y}_i$ for the target video. Detailed architecture is shown in Appendix \fcolorbox{red}{white}{F.2}.

\paragraph{Training Pipeline and Data Leakage Prevention.} Our model is trained end-to-end. We use a combined loss $L = L_{pred} + \gamma L_{mhp}$, where $L_{pred}$ is the nMSE regression loss for the final popularity prediction, and $L_{mhp}$ is the negative log-likelihood of the MHP module (as described in Sec.~\ref{sec:theorem1}). To strictly prevent data leakage: (1) During training, when retrieving neighbors for a target video $V_i$, $V_i$ itself is explicitly excluded from the retrieval candidate pool. (2) At test time, the retrieval module queries only the memory bank built from the training set. No features or labels from the test set are present in the memory bank during inference. Thus, the ground-truth labels $Y_j$ used in Eq.~\ref{eq:H_i} are always from training samples.

\begin{figure}[tbp]
\centering
\includegraphics[width=\columnwidth]{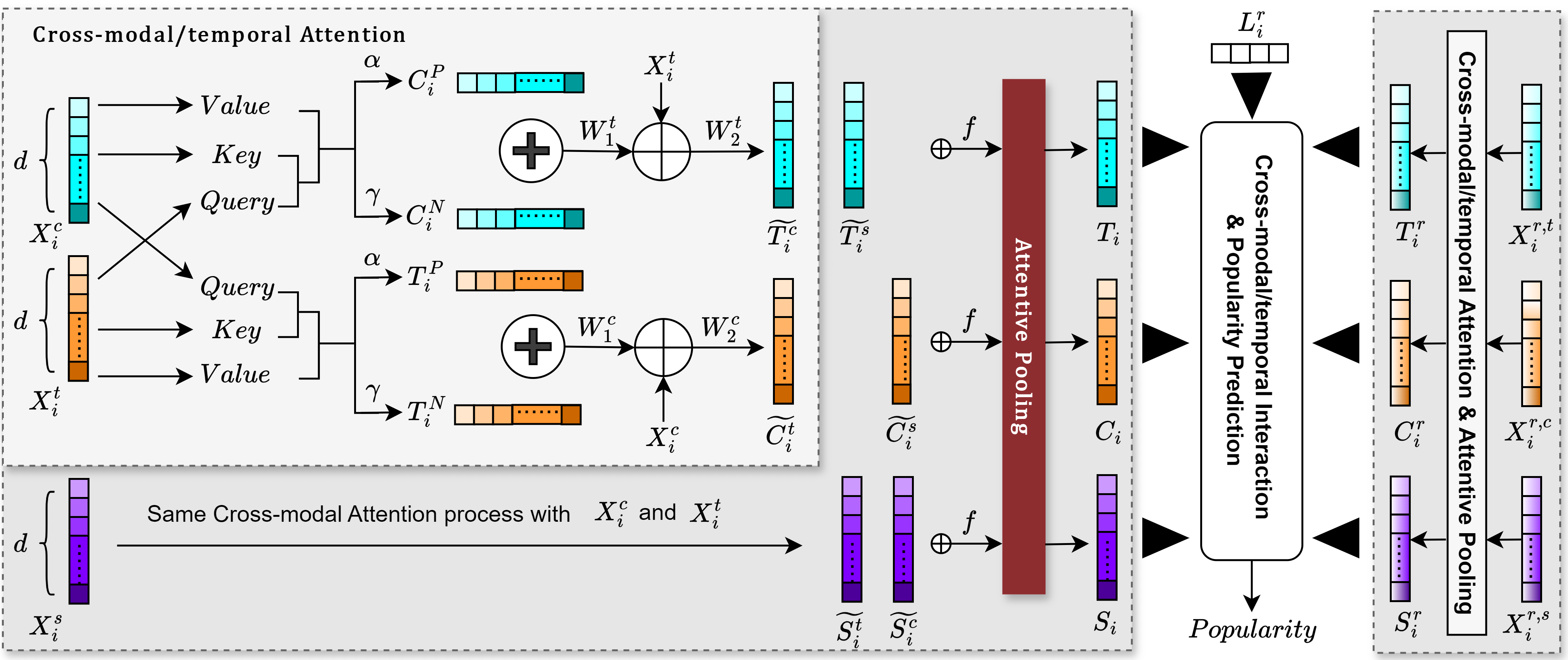}
\caption{Cross-modal/temporal Attention and Interaction with Retrieval Enhancement. Feature embeddings of both the target video and retrieved videos are integrated through cross-modal attention and interaction, which finally generate the comprehensive popularity prediction.}
\Description{A detailed interaction-and-prediction architecture. On the left, audio-visual, textual, and temporal features of the target video enter pairwise cross-modal and temporal attention modules. Query, key, and value connections are drawn between different feature streams, and the resulting aligned representations are combined and reduced through
attentive pooling. Retrieved videos undergo the same cross-modal and temporal alignment process in parallel. The pooled target representations are then compared with the aligned retrieved representations through cross-sample interactions, while retrieved popularity labels provide an additional information source. The target features, retrieved features, label information, and interaction features are finally concatenated and passed through a prediction network to produce the popularity estimate.}
\label{fig6}
\end{figure}

\section{Experiments}
\subsection{Datasets}
To evaluate the effectiveness of M$^3$TR, we conduct experiments on the micro-video dataset: \textbf{MicroLens-100k} ~\cite{ML100}. Another dataset we employ is the Tiktok Micro-video training-set which is provided by the \textbf{2024 INFORMS Data Challenge} committee. While the public INFORMS 2024 dataset only provides a single timestamp per video, our participation in the challenge granted us access to the full event-level user interaction sequences (e.g., likes, comments, shares with timestamps), which are used in this study for our Mamba-Hawkes Process module. The detailed information of these two datasets are listed in Table~\ref{dataset-info}and other existing datasets that we thoroughly reviewed are in Appendix \fcolorbox{red}{white}{J}.

\begin{table}[h!]
  \begin{center}
    \resizebox{\columnwidth}{!}{%
    \begin{tabular}{cccccc} 
    \toprule
      \textbf{Dataset} & \textbf{\#Video} & \textbf{\#User} & \textbf{\#Train} & \textbf{\#Val} & \textbf{\#Test}\\
      \hline
      MicroLens-100k & $19,724^{\ast}$ & $100,000$ & $15779$ & $1973$ & $1972$\\ 
      2024 INFORMS Challenge & 2200 & 8625 & 1760 & 220 & 220 \\
      \bottomrule
    \end{tabular}
    }
    $\ast $: 14 videos are disposed due to file corruption.
  \end{center}
\caption{Statistics of dataset.}
\label{dataset-info}
\end{table}

\subsection{Implementation}
\subsubsection{Evaluation Metrics}
We adopt the normalized Mean Square Error (nMSE) as the main parameter to measure the performance, and report Spearman's Rank Correlation (SRC) coefficient, Pearson linear correlation coefficient (PLCC), and Mean Absolute Error (MAE). Detailed metric definition is displayed in Appendix \fcolorbox{red}{white}{K}.

\subsubsection{Experimental Settings}
All our experiments are conducted on a Linux server with 4 RTX 3090 GPUs.
More detailed experimental settings are provided in Appendix \fcolorbox{red}{white}{L}.

\subsubsection{Hyper-Parameter Settings}
We adopt the Bayesian Optimization~\cite{Bayes}, searching through the parameter hyper-space, and find the lowest nMSE when setting [learning\_rate, batch\_size, weight\_decay, dropout, $\alpha$, K]=[1e-5, 64, 0.001, 0, 0.8, 10]. We apply the above set of parameters to acquire our result and compare M$^3$TR with other baselines.
\subsection{Performance Comparison}
\subsubsection{Baseline Models}
To evaluate our model, we conduct experiments with 9 competitive baselines of different methods, including SVR ~\cite{SVR}, HyFea ~\cite{HyFea}, CLSTM ~\cite{CLSTM}, TMALL ~\cite{TMALL}, MASSL ~\cite{massl}, CBAN ~\cite{CBAN}, HMMVED~\cite{HMMVED}, MTFM~\cite{zhao2024mftm} and MMRA ~\cite{MMRA}. Detailed information is summarized in Appendix \fcolorbox{red}{white}{I}.

\subsubsection{Performance}
The performance of baseline models and our model on the dataset is presented in Table \ref{tab:performance}. The results demonstrate that M$^3$TR outperforms all baselines .\par
\begin{figure*}
    \centering
    \includegraphics[width=0.95\linewidth]{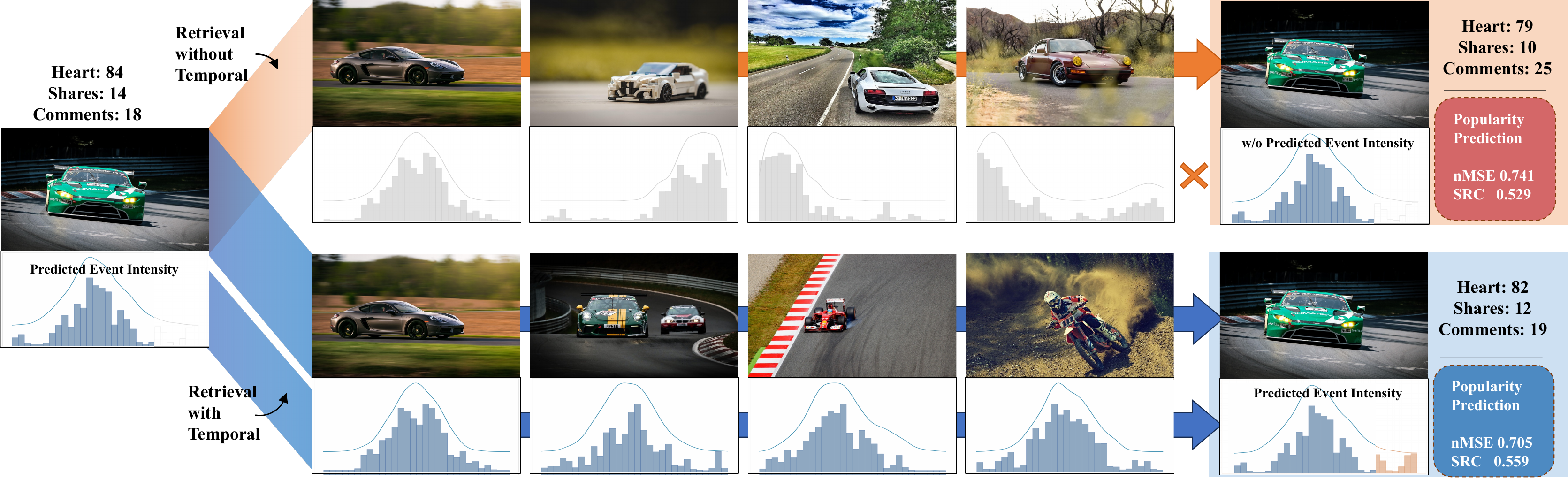}
    \caption{Qualitative experimental test results comparing traditional content-only retrieval with M$^{3}$TR temporal-aware retrieval. The target video (left) has a ground truth popularity, and its predicted event intensity follows a distinct bell curve.~\textbf{Top Row (Retrieval without Temporal):} The conventional content-only method retrieves thematically similar videos (other cars). However, their popularity trajectories are flat and dissimilar, offering poor predictive signals. This reliance on irrelevant patterns leads to an inaccurate prediction and poor experimental test metrics.~\textbf{Bottom Row (Retrieval with Temporal):} In contrast, our $M^{3}TR$ framework retrieves videos that share a similar \textit{temporal DNA}---a matching bell-shaped intensity curve---even if the content is not identical.}
    \Description{A qualitative retrieval example arranged in two horizontal branches. On the far left, the target car video has ground-truth engagement values of 84 hearts, 14 shares, and 18 comments, together with a bell-shaped event-intensity pattern. The upper branch represents content-only retrieval and contains four visually similar car videos. Their associated temporal patterns differ substantially from the target pattern. This branch predicts 79 hearts, 10 shares, and 25 comments, with nMSE 0.741 and SRC 0.529. The lower branch represents temporal-aware retrieval and contains four videos whose visual content is not necessarily identical but whose event-intensity patterns more closely resemble the target's bell-shaped pattern. This branch predicts 82 hearts, 12 shares, and 19 comments, with nMSE 0.705 and SRC 0.559.}
    \label{fig:placeholder}
\end{figure*}

\begin{table*}[h]
\centering
\begin{tabular}{c|cccc|cccc}
    \toprule
    \multirow{2}{*}{Model} & 
    \multicolumn{4}{c|}{\textbf{MicroLens-100k}} & 
    \multicolumn{4}{c}{\textbf{2024 INFORMS Data Challenge Dataset}} \\
    & nMSE$(\downarrow)$ & SRC$(\uparrow)$ & MAE$(\downarrow)$ & PLCC$(\uparrow)$ 
    & nMSE$(\downarrow)$ & SRC$(\uparrow)$ & MAE$(\downarrow)$ & PLCC$(\uparrow)$ \\
    \midrule
    CBAN     & 0.817\eb{0.021} & 0.466\eb{0.023} & 25.20\eb{1.29} & 0.429\eb{0.018} & 0.941\eb{0.051} & 0.221\eb{0.014} & 39.03\eb{1.85} & 0.350\eb{0.018} \\
    CLSTM    & 0.842\eb{0.032} & 0.471\eb{0.021} & 25.53\eb{1.41} & 0.421\eb{0.024} & 0.936\eb{0.049} & 0.316\eb{0.018} & 42.17\eb{2.71} & 0.316\eb{0.019} \\
    HMMVED   & 0.893\eb{0.040} & 0.353\eb{0.013} & 27.40\eb{1.32} & 0.323\eb{0.014} & 0.951\eb{0.042} & 0.262\eb{0.014} & 45.63\eb{2.28} & 0.199\eb{0.014} \\
    MASSL    & 0.927\eb{0.046} & 0.380\eb{0.026} & 25.16\eb{1.26} & 0.329\eb{0.014} & 1.004\eb{0.059} & 0.251\eb{0.017} & 40.90\eb{2.25} & 0.251\eb{0.014} \\
    SVR      & 1.019\eb{0.051} & 0.405\eb{0.015} & 25.04\eb{1.40} & 0.321\eb{0.013} & 1.203\eb{0.066} & 0.285\eb{0.018} & 45.45\eb{2.77} & 0.229\eb{0.013} \\
    MFTM     & 0.876\eb{0.031} & 0.414\eb{0.011} & 26.42\eb{1.37} & 0.352\eb{0.025} & 0.990\eb{0.054} & 0.301\eb{0.021} & 44.66\eb{2.93} & 0.292\eb{0.019} \\
    MMRA     & 0.785\eb{0.024} & 0.495\eb{0.025} & 24.34\eb{1.21} & 0.480\eb{0.020} & 0.911\eb{0.050} & 0.387\eb{0.026} & 39.21\eb{2.26} & 0.373\eb{0.025} \\
    Hyfea    & 0.840\eb{0.039} & 0.432\eb{0.022} & 26.10\eb{1.38} & 0.408\eb{0.022} & 0.953\eb{0.058} & 0.245\eb{0.019} & 36.81\eb{2.84} & 0.284\eb{0.016} \\
    TMALL    & 0.856\eb{0.033} & 0.440\eb{0.022} & 26.11\eb{1.22} & 0.401\eb{0.028} & 0.913\eb{0.042} & 0.296\eb{0.021} & 40.20\eb{2.45} & 0.345\eb{0.020} \\
    \midrule
    \rowcolor{gray!20}
    \textbf{M$^3$TR} & \textbf{0.705}\eb{0.020} & \textbf{0.559}\eb{0.024} & \textbf{23.10}\eb{1.01} & \textbf{0.557}\eb{0.025} 
                      & \textbf{0.854}\eb{0.037} & \textbf{0.433}\eb{0.020} & \textbf{31.64}\eb{1.59} & \textbf{0.406}\eb{0.016} \\
    \bottomrule
\end{tabular}
\caption{Performance Comparison of Baselines and M$^3$TR. We report the average results with standard deviations across 3 independent runs with different random seeds. Our proposed $M^3$TR consistently outperforms all competitive baselines across both datasets. Notably, on the MicroLens-100k dataset, $M^3$TR achieves a significant reduction in nMSE and a substantial gain in SRC demonstrating its superior ability to capture complex micro-video popularity dynamics and maintain ranking consistency.}
\label{tab:performance}
\end{table*}

\subsection{Ablation Study}
\subsubsection{Qualitative Analysis}
\begin{figure}[htbp]
    \centering
    \includegraphics[width=\linewidth]{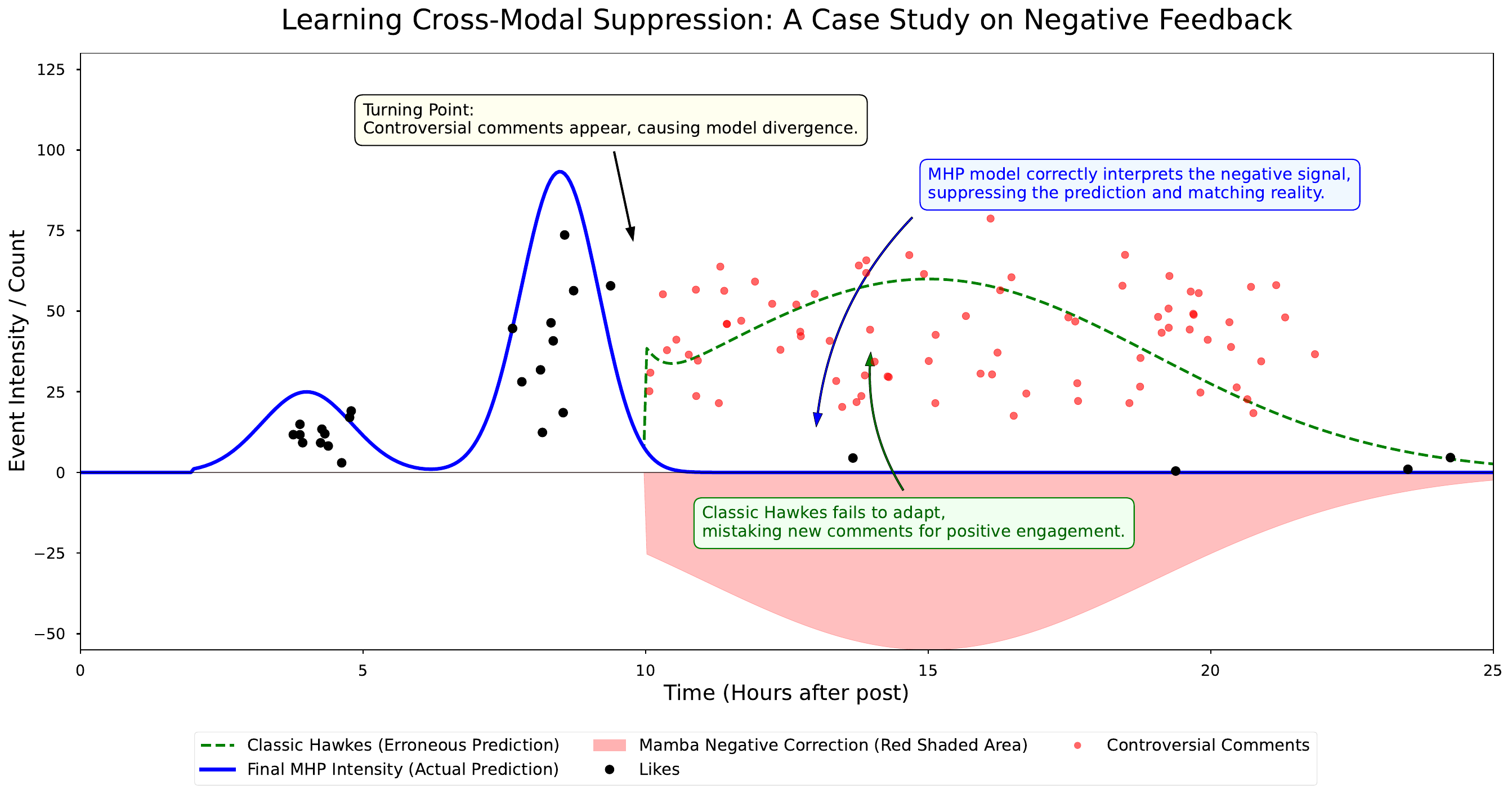}
    \caption{A case study visualizing the internal dynamics of the MHP module, demonstrating its ability to handle complex, cross-modal signals. This chart contrasts two predictive models tracking a video's engagement. Initially, both the `\textcolor{blue}{Final MHP Intensity}' (blue solid line) and the `\textcolor{darkgreen}{Classic Hawkes}' model (green dashed line) observe a series of `\textcolor{black}{Likes}' (black dots). However, after the 10-hour mark, a new cross-modal event occurs: a wave of `\textcolor{red}{Controversial Comments}' (red dots). The `\textcolor{darkgreen}{Classic Hawkes}' process, which is blind to this new temporal signal , fails to adapt and erroneously predicts continued engagement growth. In sharp contrast, our Mamba component detects the suppressive nature of these comments and introduces a `\textcolor{red}{Mamba Negative Correction}' (red shaded area). This crucial correction allows our `\textcolor{blue}{Final MHP Intensity}' (blue solid line) to accurately model the real-world outcome.}
    \Description{A line chart covering approximately 25 hours after a video is posted. Likes are concentrated in the early part of the timeline, while controversial comments begin around hour 10 and continue through the later period. Before the turning point, the classical Hawkes curve and the final MHP curve both respond to bursts of positive interaction. After the comments appear, the classical Hawkes curve rises into a broad positive peak, whereas the MHP curve is suppressed and returns toward a low intensity. A shaded negative region below the curves shows the correction contributed by the Mamba component; this region becomes large after the comments begin and gradually decreases later. Annotations identify the onset of the controversial comments and the subsequent divergence between the two predictions.}
    \label{fig:mhp_case_study}
\end{figure}

The importance of our Mamba-Hawkes Process (MHP) module is best illustrated through a qualitative case study. Figure \ref{fig:mhp_case_study} visualizes a complex scenario where an initial surge of positive interactions (Likes) is followed by a wave of controversial Comments. The classic Hawkes process (green dashed line), only observing the self-exciting nature of likes, erroneously predicts a continued, explosive growth. In contrast, our Mamba component learns the complex, cross-modal dependency and provides a crucial negative correction (red shaded area), recognizing that controversial comments suppress future like activity. The final MHP intensity (blue solid line) accurately models the resulting stall in engagement. This case study demonstrates the MHP's unique ability to capture nuanced, non-linear dynamics that simpler temporal models miss. A more detailed breakdown of this case study is in Appendix \fcolorbox{red}{white}{M.1}. And we present a detailed qualitative case study across various popularity archetypes in Appendix \fcolorbox{red}{white}{M.2} to further provide an intuitive understanding of these dynamics.
\subsubsection{Quantitative Results}
The comprehensive quantitative results of ablation experiments corroborate these qualitative findings across three key dimensions: overall architecture robustness, temporal modeling strategies, and multi-modal integration. 

First, the four primary variants (Table \ref{tab:main_ablation}) show substantial performance drops for \textbf{noTemporal} and \textbf{noMHP}, highlighting the indispensable role of temporal dynamics. The full M$^3$TR model achieves the best overall performance and produces the most accurate popularity predictions for both the stratified Top and Bottom groups, demonstrating its robustness across highly popular (head) and less popular (tail) videos.

\begin{table}[htbp]
    \centering
    \setlength{\tabcolsep}{3pt} 
    \resizebox{0.8\columnwidth}{!}{
    \begin{tabular}{lcccc}
        \toprule
        \textbf{Prediction} & \textbf{noTemp} & \textbf{noRet} & \textbf{noMHP} & \textbf{M$^3$TR} \\
        \midrule
        $Top50$ & 75.000 & 77.908 & 71.688 & \textbf{88.030} \\
        $Top200$ & 57.690 & 56.513 & 55.616 & \textbf{64.266} \\
        $Top500$ & 43.111 & 44.420 & 41.604 & \textbf{44.443} \\
        $Bottom500$ & 27.714 & 26.017 & 26.339 & \textbf{25.866} \\
        $Bottom200$ & 22.638 & 22.544 & 22.818 & \textbf{22.447} \\
        $Bottom50$ & 23.409 & 22.414 & \textbf{19.536} & 22.183 \\
        \midrule
        nMSE($\downarrow$) & 0.7413 & 0.7980 & 0.8002 & \textbf{0.7051} \\
        SRC($\uparrow$) & 0.5286 & 0.5047 & 0.5064 & \textbf{0.5587} \\
        MAE($\downarrow$) & 23.814 & 25.198 & 24.359 & \textbf{23.097} \\
        \bottomrule
    \end{tabular}
    }
    \caption{Ablation study results demonstrating the effectiveness of the proposed components in M$^{3}$TR. We compare four primary variants (no Temporal Information, no Retrieval, no Mamba Hawkes Process, and full M$^{3}$TR) evaluated across different popularity rank groupings (Top/Bottom) and overall metrics. The results indicate that our full framework yields the most robust predictions across both highly popular and less popular videos.}
    \label{tab:main_ablation}
\end{table}

Second, the comparison of temporal modeling strategies (Table \ref{tab:temporal_ablation}) further reveals \textit{how} these dynamics should be captured. Simple approaches such as Padding\&Masking and the Poisson Process are insufficient for effectively modeling temporal information. Although the Neural Hawkes Process improves performance, our Mamba Hawkes Process (MHP) achieves the best results. By explicitly modeling the mutually exciting and decaying patterns of popularity dynamics, MHP captures nuanced event-driven cascades and produces a more faithful temporal representation.

Finally, the modality ablation (Table \ref{tab:modality_ablation},), which separately examines retrieved labels, visual content, audio, and text, confirms the value of multi-modal fusion. Each modality contributes positively to predictive accuracy. Notably, even without retrieved labels, the model maintains strong performance and continues to outperform the baselines. These results demonstrate that integrating temporal event trajectories with multi-modal video content provides a robust foundation for popularity prediction.

\begin{table}[t]
    \centering

    \begin{subtable}[t]{0.49\columnwidth}
        \centering
        \resizebox{\linewidth}{!}{
        \begin{tabular}{lccc}
            \toprule
            \textbf{Temporal} & \textbf{nMSE} & \textbf{SRC} & \textbf{MAE} \\
            \midrule
            None & 0.741 & 0.529 & 23.81 \\
            Pad\&Mask & 0.800 & 0.506 & 24.36 \\
            Poisson & 0.765 & 0.525 & 24.02 \\
            Neural HP & 0.723 & 0.537 & 23.44 \\
            \rowcolor{gray!20}
            Mamba HP & \textbf{0.705} & \textbf{0.559} & \textbf{23.10} \\
            \bottomrule
        \end{tabular}
        }
        \caption{Comparison of different temporal modeling strategies.}
        \label{tab:temporal_ablation}
    \end{subtable}
    \hfill
    \begin{subtable}[t]{0.49\columnwidth}
        \centering
        \resizebox{\linewidth}{!}{
        \begin{tabular}{lccc}
            \toprule
            \textbf{Ablation} & \textbf{nMSE} & \textbf{SRC} & \textbf{MAE} \\
            \midrule
            w/o label & 0.746 & 0.506 & 24.76 \\
            w/o audio & 0.739 & 0.525 & 23.53 \\
            w/o visual & 0.735 & 0.519 & 23.49 \\
            w/o text & 0.718 & 0.534 & 23.37 \\
            \rowcolor{gray!20}
            with all & \textbf{0.705} & \textbf{0.559} & \textbf{23.10} \\
            \bottomrule
        \end{tabular}
        }
        \caption{Ablation of different input modalities.}
        \label{tab:modality_ablation}
    \end{subtable}

    \caption{Ablation studies of temporal modeling strategies and input modalities. Left: Mamba Hawkes Process achieves the best performance among the temporal modeling strategies. Right: each modality contributes positively to the overall predictive performance.}
    \label{tab:sub_ablations}
\end{table}

As shown in Table~\ref{tab:main_ablation}, the predicted popularity scores for the six groups followed the order: Top50 $>$ Top100 $>$ Top200 $>$ Bottom200 $>$ Bottom100 $>$ Bottom50, indicating that the prediction model performs as expected. Furthermore, compared to the actual number of comments in each group, M$^3$TR achieved the best performance in most cases. As for the evaluation metrics, M$^3$TR outperformed the other three variants in nMSE, SRC, and MAE, validating the effectiveness of incorporating historical and audio information in the retrieval stage.\par

\subsection{Complexity Discussion and Practicality}\label{sec:complexity}
Our architecture follows a two-stage design: a one-time, computationally intensive \textbf{offline stage} for constructing a retrieval-ready memory bank, and an efficient \textbf{online stage} for real-time inference.

\textbf{Offline Memory Bank Construction.}
This stage processes each video in the historical corpus through several computationally intensive operations, including multi-modal feature extraction (i.e., ViT and AST), MHP-based temporal embedding generation, and unified retrieval vector construction (i.e., BLIP-2 and CLAP). To quantify the computational cost, we processed nearly 20,000 samples using four RTX 3090 GPUs. Visual and acoustic captioning required 398 and 182 minutes, respectively. Temporal embedding generation took 13 minutes, while textual, visual, and acoustic embedding extraction required 8, 33, and 146 minutes, respectively. Finally, UAE generation took only 5 minutes. These features are computed once for each video and stored in an efficient Approximate Nearest Neighbor (ANN) search index. For our experiments, we utilized FAISS with a [e.g., IVF1024, PQ64] index~\cite{faiss_library_2024,johnson2019billion,jegou2011product}, which balances search speed and memory. The total cost scales linearly with video number and can be distributed across a compute cluster. Once the offline pre-processing is complete, the entire model is trained in merely 24 minutes.

\textbf{Online Real-Time Inference.}
For a new target video, the trained model supports low-latency, real-time prediction. After extracting its features, we employ Approximate Nearest Neighbor search, e.g., FAISS~\cite{faiss}, reducing retrieval complexity to a scalable $O(\log N_{\text{bank}})$. The final fusion module then operates only on the retrieved set of $S$ exemplars, resulting in a low complexity of approximately $O(S \cdot d^2)$ that is independent of the memory bank size. This two-stage design therefore enables efficient and scalable deployment in real-world applications. A detailed latency breakdown of the online components is provided in Appendix~\ref{appendix:complexity_deployment}.

\section{Conclusion}
In this work, we introduce M$^3$TR, a framework that establishes a new SOTA in MVPP. Our central contribution is the insight that a video's underlying temporal archetype is as critical as its content for long-term forecasting. We address the limitations of prior art by synergizing two core innovations: (1) a Mamba-Hawkes Process that captures the intricate, self-exciting nature of user interactions, and (2) a temporal-aware retrieval mechanism that identifies historical videos based on both content and dynamic patterns. Our results validate that retrieving videos with similar popularity trajectories provides more robust signals for forecasting than content similarity alone. We believe this paradigm of temporal-aware retrieval can be generalized to other user-generated content platforms, and offers a new conceptual insight for future research on dynamic content popularity.

\clearpage
\begin{acks}
This work was supported in part by the National Natural
Science Foundation of China (NO. 62472284), Shanghai
Key Laboratory of Scalable Computing and Systems, SJTU
Kunpeng \& Ascend Center of Excellence, and Shanghai
Jiao Tong University under “Bo Le” Grant AO120001/074.
\end{acks}
\bibliographystyle{ACM-Reference-Format}
\balance
\bibliography{references}

\clearpage
\appendix
\setcounter{equation}{0}
\setcounter{figure}{0}

\begin{center}
    \Large\bfseries Appendix \\
\end{center}

\section{More Challenges Faced by Previous Methods}
\label{challenges}
\begin{figure*}[ht]
    \centering
    \begin{subfigure}{0.4\textwidth}
        \includegraphics[width=\linewidth]{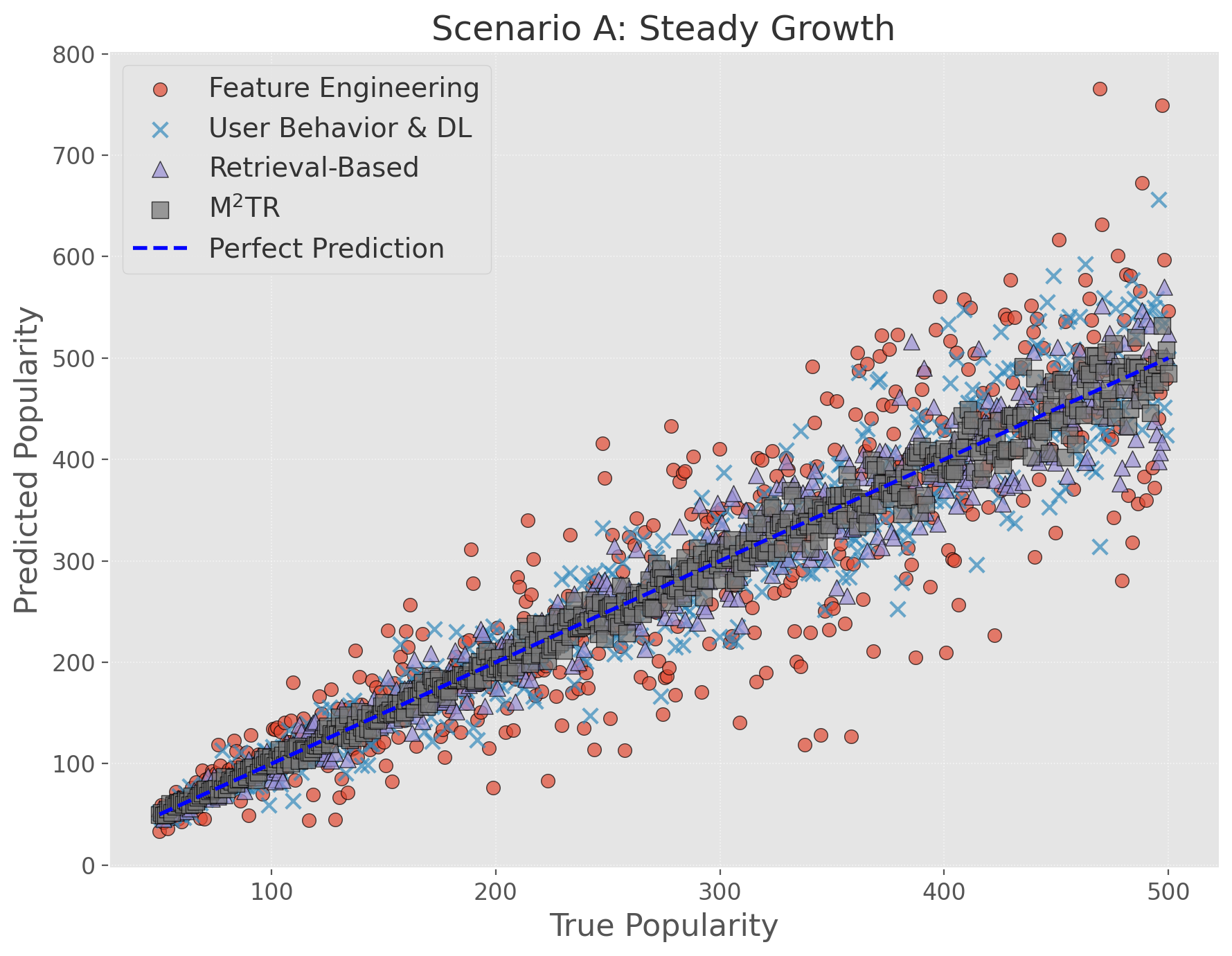}
        \caption{Scenario A: Steady Growth.}
        \label{A}
    \end{subfigure}%
    \hfill
    \begin{subfigure}{0.4\textwidth}
        \includegraphics[width=\linewidth]{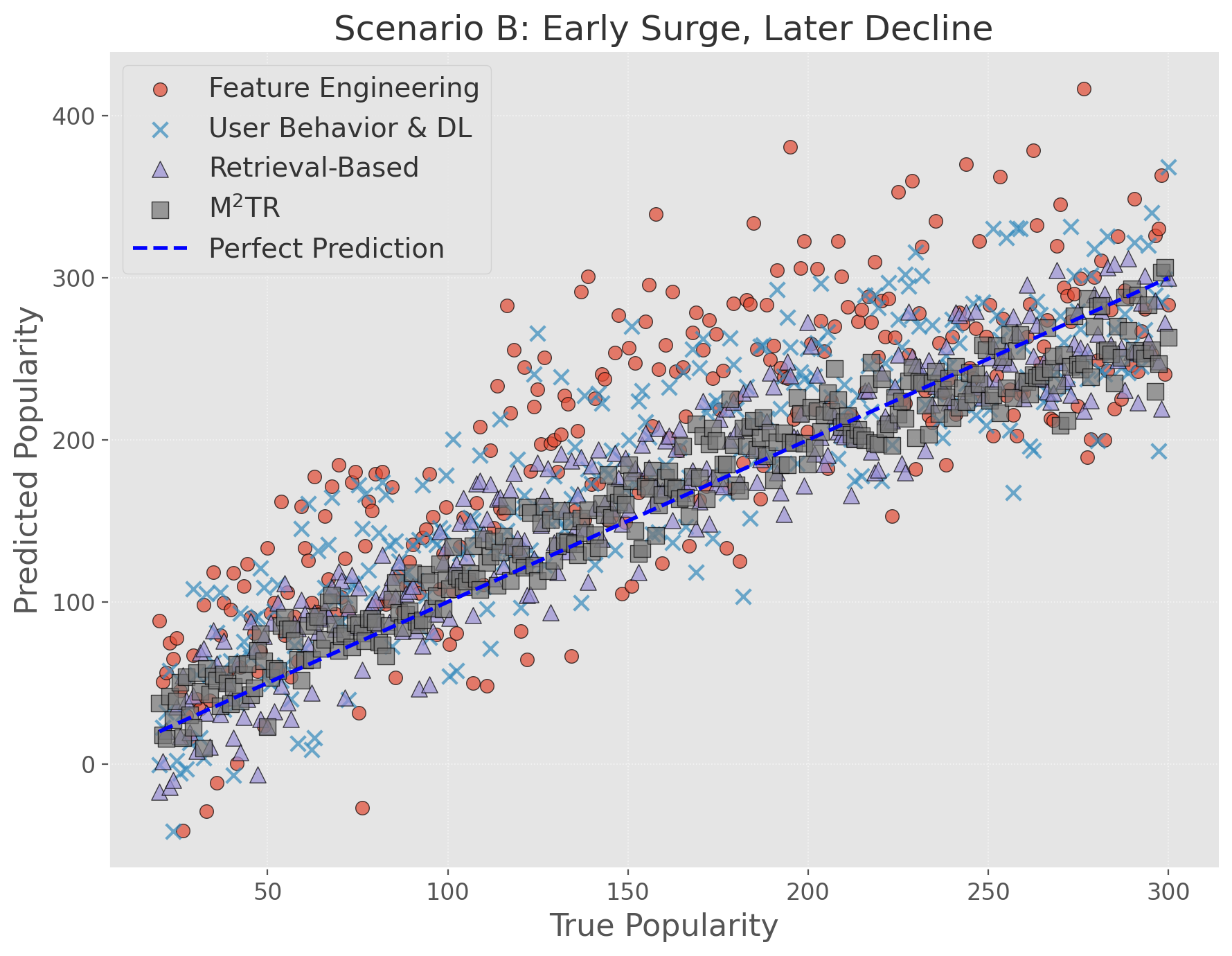}
        \caption{Scenario B: Early Surge, Later Decline.}
        \label{B}
    \end{subfigure}%
    \hfill
    \begin{subfigure}{0.4\textwidth}
        \includegraphics[width=\linewidth]{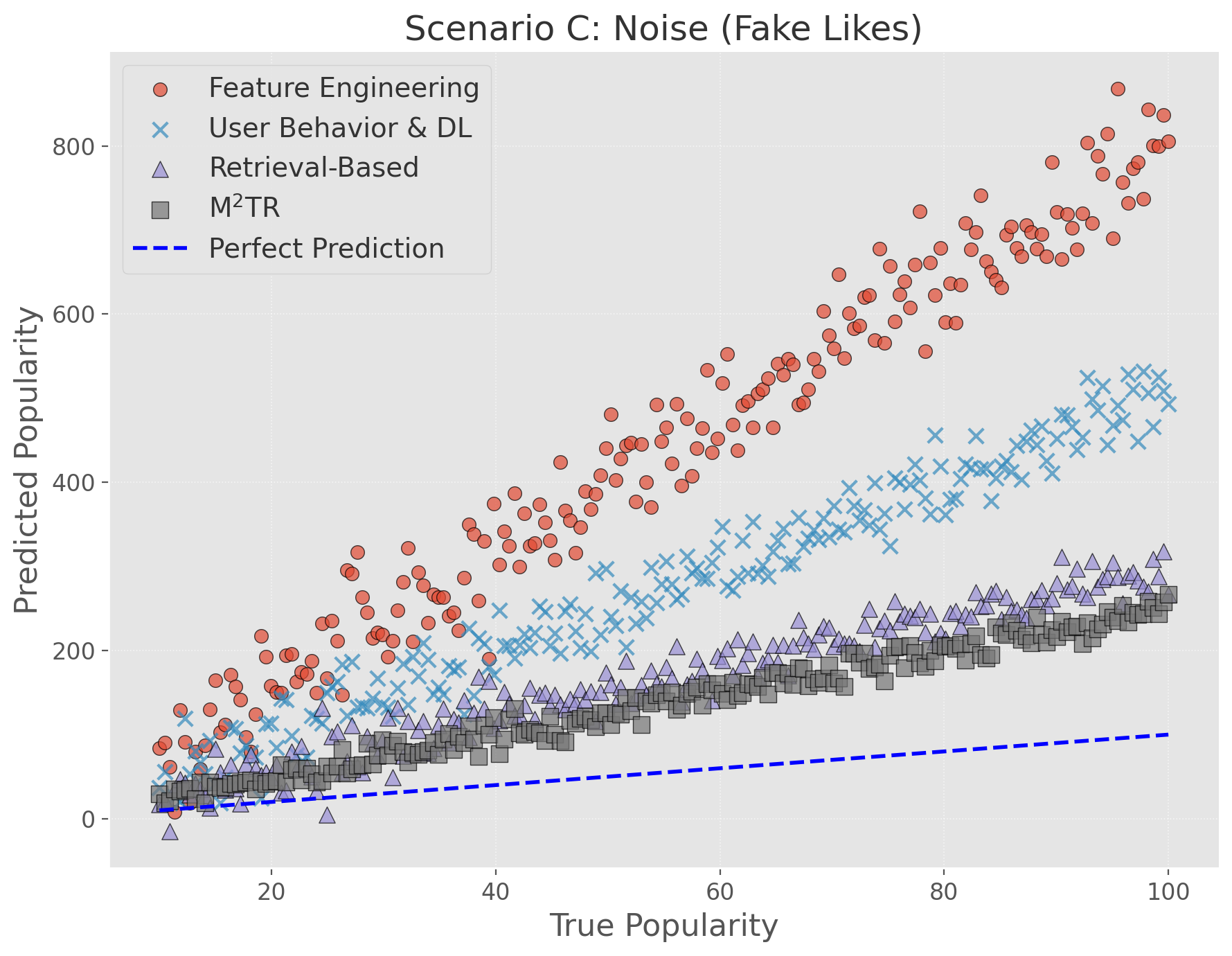}
        \caption{Scenario C: Noise Interfere}
        \label{C}
    \end{subfigure}
    \caption{Comparisons of Different Scenarios of Video Popularity Trends. (a) Steady growth, (b) Early surge with later decline, (c) Noise from fake likes.}
    \Description{
Three scatter plots compare predicted popularity with true popularity
under different conditions. Each plot contains results from feature
engineering, user-behavior or deep-learning models, retrieval-based
models, and M3TR, together with a dashed diagonal line representing
perfect prediction. In the steady-growth scenario, the predictions
generally follow the diagonal, although the methods show different
amounts of dispersion. In the early-surge and later-decline scenario,
the scatter around the diagonal becomes wider. In the fake-like noise
scenario, the methods separate strongly: feature-engineering
predictions show the largest upward deviation from the ideal line,
user-behavior models show a smaller but still substantial deviation,
and retrieval-based methods and M3TR remain closer to the lower
prediction range.
}
    \label{fig:scenarios}
\end{figure*}

Current micro-video popularity prediction research suffers from two main limitations, demonstrable through analysis of the MicroLens-50k dataset~\cite{ML100} and visualized in Figures~\ref{fig:scenarios}. First, existing models lack a comprehensive understanding of user interests. While some approaches focus on temporal user feedback (likes, comments, shares, etc.), few effectively integrate these with retrieval mechanisms. This prevents models from capturing contextualized user interests. For example, retrieval methods relying solely on user behavior similarity often fail to distinguish between videos with similar initial popularity but diverging later trends, as illustrated in Figure~\ref{B} (the ``initial rise, then decline" scenario). These methods don't adequately connect content and dynamic feedback (e.g., increasing negative comments) to accurately reflect the changing user interest. Furthermore, fine-grained interest understanding, like differentiating user preferences for various video modalities, is often neglected.

Second, existing models, including retrieval-based ones, show deficiencies in accuracy, robustness, cold-start handling, and adaptability. Insufficient information use and limitations in retrieval (e.g., susceptibility to noise, as shown in Figure~\ref{C} for the ``noise interference" scenario) lead to prediction biases. While retrieval-based models generally outperform feature-engineering and user-behavior/deep-learning models in scenarios with consistent positive feedback (Figure~\ref{C}, ``positive feedback" scenario), analysis of MicroLens-50k data reveals that all model types struggle with contextualized interests, noise, and cold-start situations (sparse early feedback). This is evident across all scenarios, including the steady growth scenario in Figure~\ref{A}, where even the best-performing retrieval-based models show some deviation from perfect prediction. This highlights the critical need for a more effective integration of multi-modal information and temporal user feedback within the retrieval framework to improve prediction accuracy and robustness, especially in more complex situations like those depicted in Figures~\ref{B} and~\ref{C}.

\begin{figure}
    \centering
    \includegraphics[width=\columnwidth]{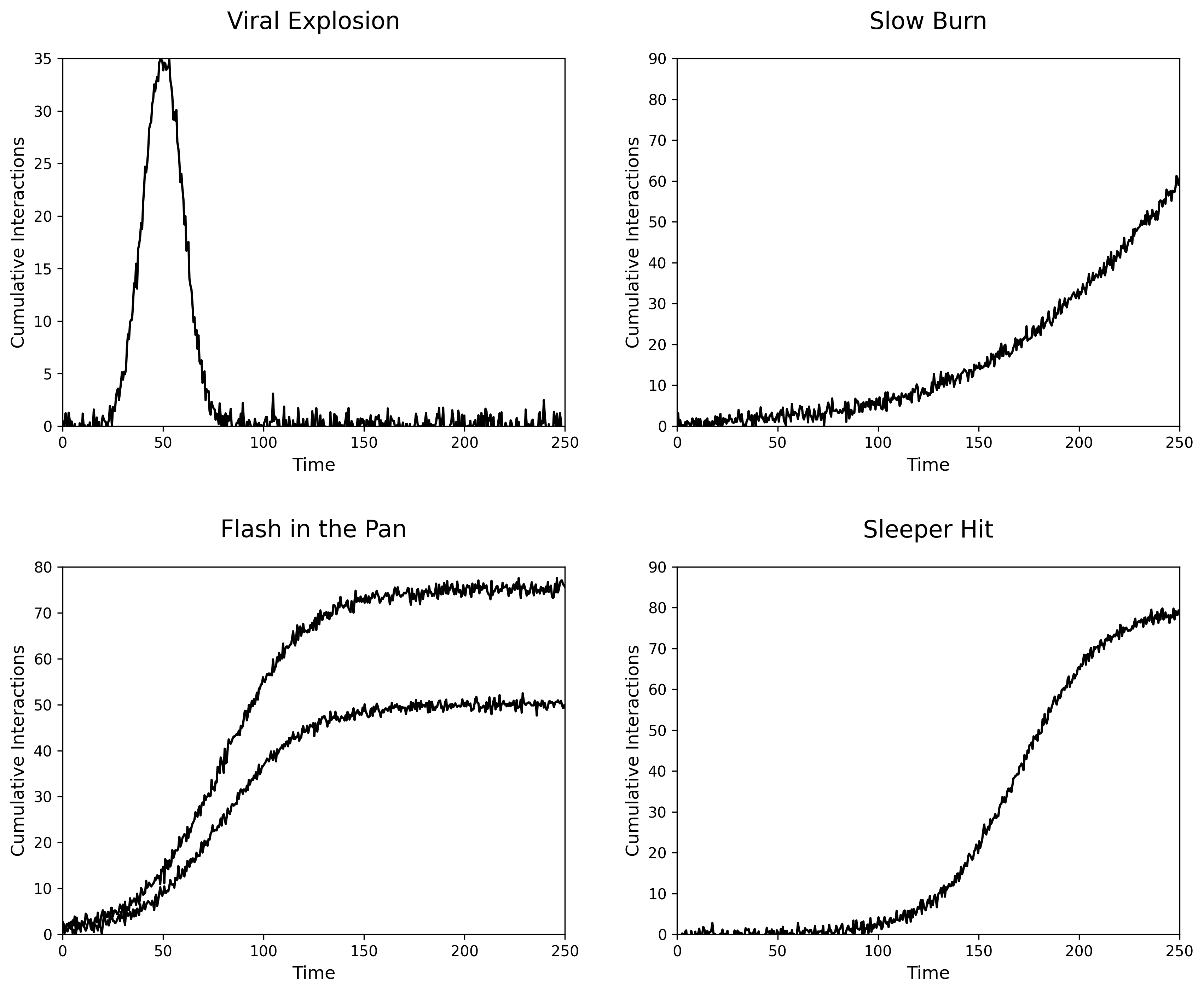}
    \caption{Diversity of Micro-Video Popularity Trajectories}
    \Description{
Four small time-series plots illustrate visually distinct popularity
patterns. The upper-left plot, labeled Viral Explosion, contains a
sharp and narrow burst of activity early in the timeline followed by
a rapid decline. The upper-right Slow Burn plot rises gradually across
most of the observation period. The lower-left Flash in the Pan plot
shows rapid early growth followed by flattening into a plateau rather
than continued acceleration. The lower-right Sleeper Hit plot remains
near a low level for an extended initial period and then undergoes a
delayed, steep rise before saturating. The four panels emphasize the
substantial variation in the timing, slope, and persistence of
micro-video popularity dynamics.
}
    \label{fig:appendix_trajectories}
\end{figure}

To contextualize the challenges inherent in Micro-Video Popularity Prediction (MVPP), it is crucial to first appreciate the sheer diversity of engagement trajectories that videos exhibit over their lifecycle. The assumption of a single, uniform growth model is untenable, as different videos possess fundamentally distinct temporal signatures. A predictive model that fails to account for this diversity is bound to have limited real-world applicability. Figure~\ref{fig:appendix_trajectories} provides a qualitative illustration of four common yet disparate popularity archetypes, underscoring the necessity for a sophisticated model capable of understanding and forecasting these complex temporal dynamics.

``Viral Explosion" is characterized by a steep, exponential accumulation of engagement within a very short timeframe. These videos achieve peak popularity rapidly before the growth rate saturates. ``Slow Burn" exhibits a more gradual but sustained growth in popularity. These videos accumulate engagement at a steady, often linear, rate over a much longer period, lacking the initial explosive phase. ``Flash in the Pan" (Early Surge, Later Decline) represents a particularly challenging scenario where a video shows strong initial promise but fails to maintain momentum. User interest or relevance wanes, leading to a premature plateau or even a decline in the rate of new engagement. And ``Sleeper Hit" represents that these videos remain dormant with minimal engagement for an initial period before a delayed catalyst (e.g., a mention and a change in algorithm) triggers a sudden and often rapid rise in popularity.

The existence of these disparate ``temporal DNAs" is the central challenge that our work addresses. Two videos with nearly identical multi-modal content features could follow entirely different paths; for example, one might become a ``Viral Explosion" while the other is a ``Flash in the Pan". This observation strongly motivates our core hypothesis: a successful prediction framework must look beyond static content similarity and develop a deep, nuanced understanding of a video's temporal signature. The M$^3$TR framework, particularly with its Mamba-Hawkes Process (MHP) module, is specifically designed to model these intricate event dynamics. Furthermore, our temporal-aware retrieval mechanism is engineered to identify historically relevant examples that share a similar temporal DNA, providing the crucial context that content-only models invariably miss.

\section{Mamba Architecture}
\label{app:architecture}
We begin with the continuous form of the classic State Space Models (SSM), gradually introducing how to discretize them, implement input-dependent dynamics, and optimize for parallel scan on GPUs for efficient training and inference.

\subsection{Intro: From Traditional Transformers to Mamba}
Transformers and their self-attention mechanism have shown exceptional performance across various sequence tasks, such as natural language, genomics, and audio signals. However, the self-attention module in Transformer models exhibits a quadratic time and memory complexity with respect to the sequence length \(L\), making them prohibitively expensive for long sequences. Several alternatives to address this inefficiency have been proposed, including:
\begin{itemize}
    \item \textbf{Linear Attention:} Using kernel methods or approximations to reduce the attention complexity to linear, though still not matching the performance of full attention in certain dense information settings like natural language modeling.
    \item \textbf{Gated Convolutions and RNN-based models:} Examples like Gated CNN and LSTM that improve local feature modeling or long-range dependencies but often underperform compared to attention models in large-scale data or complex tasks.
    \item \textbf{Structured State Space Models (SSMs):} Models like S4, S5, and H3 utilize specific structural assumptions for the state transition matrix \(\mathbf{A}\), allowing for parallelizable convolutions and efficient modeling of long sequences. However, they struggle in discrete, information-dense input domains such as text.
\end{itemize}

Mamba's introduction is driven by a core innovation: \textbf{introducing a selective mechanism into the SSM}, where the parameters of the state space model dynamically depend on the input, allowing the model to propagate or forget information based on the current input context. This design retains the SSM's ability to efficiently model long-range dependencies while addressing the shortcomings of existing methods for discrete tasks.

\subsection{Continuous-Time State Space Models Overview}

\subsubsection{Continuous Form.}
A classic (linear) state space model is described by the following continuous differential equations:

\begin{equation}
   \begin{cases}
\dot{\mathbf{h}}(t) = \mathbf{A} \mathbf{h}(t) + \mathbf{B} \mathbf{x}(t), \\
\mathbf{y}(t) = \mathbf{C} \mathbf{h}(t),
\end{cases} 
\end{equation}

\noindent where
\(\mathbf{h}(t) \in \mathbb{R}^N\) is the state vector of the system at time \(t\);
\(\mathbf{x}(t) \in \mathbb{R}\) is the input (the formulation can be extended to multi-channel inputs \(\mathbf{x}(t) \in \mathbb{R}^d\));
\(\mathbf{y}(t) \in \mathbb{R}\) is the output (again extendable to multi-channel);
\(\mathbf{A} \in \mathbb{R}^{N \times N}\), \(\mathbf{B} \in \mathbb{R}^{N \times 1}\), \(\mathbf{C} \in \mathbb{R}^{1 \times N}\) are the state transition, input projection, and output projection matrices, respectively.

Under the assumption of \textbf{Linear Time-Invariance (LTI)}, the matrices \(\mathbf{A}\), \(\mathbf{B}\), and \(\mathbf{C}\) are time-invariant. However, when extended to \textbf{Linear Time-Varying (LTV)} systems, these matrices can depend on time, enabling more flexible modeling. Mamba pushes beyond the traditional SSM by introducing input-dependent parameters, which we call \textbf{selectivity}.

\subsubsection{Discretization Process.}

In practical applications, continuous-time state space models need to be discretized. The following outlines the discretization process using Zero-Order Hold (ZOH) approximation. Assuming that \(\mathbf{x}(t)\) remains constant over the interval \([k\Delta t, (k+1)\Delta t]\), we discretize the state update equation:
\begin{equation}
   \begin{aligned}
\mathbf{h}[k+1] &= e^{\mathbf{A} \Delta t} \mathbf{h}[k] + \int_0^{\Delta t} e^{\mathbf{A}(\Delta t - \tau)} d\tau \mathbf{B} \mathbf{x}[k], \\
\mathbf{y}[k] &= \mathbf{C} \mathbf{h}[k].
\end{aligned} 
\end{equation}

By approximating the integral and simplifying, we arrive at:
\begin{equation}
    \mathbf{h}[k+1] = \bar{\mathbf{A}} \mathbf{h}[k] + \bar{\mathbf{B}} \mathbf{x}[k],
\quad
\mathbf{y}[k] = \mathbf{C} \mathbf{h}[k],
\end{equation}

\noindent where \(\bar{\mathbf{A}} = e^{\mathbf{A} \Delta t}\) and \(\bar{\mathbf{B}} = \mathbf{A}^{-1}(e^{\mathbf{A} \Delta t} - \mathbf{I}) \mathbf{B}\).

This discretized form is more suitable for digital computation, especially in sequence modeling tasks.

\subsection{From SSM to Selective SSM (SSSM)}

\subsubsection{Principle and Motivation.}
In standard SSMs (LTI), \(\bar{\mathbf{A}}, \bar{\mathbf{B}}, \mathbf{C}\) remain time-invariant, effectively making the system equivalent to a fixed convolution or linear recurrence. This property allows for efficient computation using FFT (when the sampling rate is fixed) or online recurrent scans, achieving \(O(L \log L)\) complexity. However, in tasks such as natural language modeling, where the input consists of discrete tokens, it is difficult for these models to ``selectively" retain or forget certain parts of the input sequence without introducing inefficiencies.

\textbf{Selective SSM} (SSSM) introduces input-dependent dynamics into the model, allowing the state transition matrices \(\bar{\mathbf{A}}, \bar{\mathbf{B}}, \mathbf{C}\) to vary with the current input \(\mathbf{x}[k]\):
\begin{equation}
   \begin{cases}
\mathbf{h}[k+1] = \bar{\mathbf{A}}\bigl(\mathbf{x}[k]\bigr) \mathbf{h}[k] + \bar{\mathbf{B}}\bigl(\mathbf{x}[k]\bigr) \mathbf{x}[k], \\
\mathbf{y}[k] = \mathbf{C}\bigl(\mathbf{x}[k]\bigr) \mathbf{h}[k].
\end{cases} 
\end{equation}

At each step \(k\), the model computes a set of ``selective" or ``gated" parameters based on \(\mathbf{x}[k]\), which are then used to update the state \(\mathbf{h}[k]\) and produce the output \(\mathbf{y}[k]\). This allows the model to focus on or forget particular inputs based on their relevance to the current context.

\subsubsection{Specific Forms of Input-Dependent Parameters.}
In practice, it is common to define the input-dependent parameters \(\mathbf{B}\bigl(\mathbf{x}[k]\bigr)\), \(\mathbf{C}\bigl(\mathbf{x}[k]\bigr)\), and \(\Delta\bigl(\mathbf{x}[k]\bigr)\) using a parameterized function, such as:
\begin{equation}
    \mathbf{B}\bigl(\mathbf{x}[k]\bigr) = \text{Linear}_N(\mathbf{x}[k]), \quad
\mathbf{C}\bigl(\mathbf{x}[k]\bigr) = \text{Linear}_N(\mathbf{x}[k]),
\end{equation}

\begin{equation}
    \Delta\bigl(\mathbf{x}[k]\bigr) = \text{Broadcast}_D(\text{Linear}_1(\mathbf{x}[k])).
\end{equation}

This input-dependent design makes the model more expressive, enabling selective memory and forgetting of sequence elements. The dynamic parameterization of \(\Delta\) is similar to gating mechanisms in RNNs (such as in LSTM), but with a more flexible, input-driven mechanism that adjusts how much of the previous state is retained.

\subsection{Mamba Architecture Implementation Details}

The Mamba architecture integrates the aforementioned selective SSM as its core sequence transformation block, combining it with a simplified gated MLP layer to form a unified network module. Unlike the traditional “attention + MLP” stacked blocks (e.g., Transformer blocks) or the “SSM + MLP” alternations (e.g., H3 architecture), Mamba uses a \textbf{homogeneous} design: it combines SSSM with MLP gating directly in the main branch and stacks them in each layer (or block) of the network.

The structure of the Mamba block includes:
\begin{itemize}
    \item \textbf{Projection Layer:} Linearly transforms the input channels to a higher dimension (controlled by an expansion ratio).
    \item \textbf{Non-linearity / Gating:} A multiplicative or additive gating function is applied to the higher-dimensional representation, using activations such as SiLU, Swish, or Sigmoid.
    \item \textbf{Selective SSM:} The gated representation is passed through the selective SSM block, performing sequence-wise integration and long-range modeling.
    \item \textbf{Output Projection Layer:} Projects the high-dimensional representation back to the original number of channels.
\end{itemize}

All these layers are stacked in the sequence dimension, forming the full network architecture. Empirical experiments show that this design performs exceptionally well across tasks involving text (discrete tokens), audio (continuous signals), and genomic sequences (long and discrete sequences).

\subsection{Parallel Scan and Hardware Optimization}

\subsubsection{Why Not Use FFT or Convolutions.}
In traditional LTI SSMs (e.g., S4), the state recurrence can be unfolded as a global convolution, allowing for FFT-based acceleration, achieving near \(O(L \log L)\) complexity. However, in selective SSMs, the input-dependent dynamics (\(\mathbf{A}(k), \mathbf{B}(k)\)) break the time invariance, making direct use of convolutions or FFT infeasible.

\subsubsection{Parallel Scan Algorithm and Memory Hierarchy.}
To address this issue, Mamba utilizes a hardware-friendly parallel scan algorithm to update the states in a recurrent computation mode. The key idea is:
\begin{itemize}
    \item \textbf{Chunking:} The long sequence is divided into smaller chunks, each of which can be processed in parallel on the GPU, reducing global sequential dependencies.
    \item \textbf{Prefix / Suffix Merging:} Similar to prefix sum computation, the model computes local states within each chunk first, and then merges them using a tree-like structure (in “upsweep” and “downsweep”).
    \item \textbf{Memory Management:} Most parameters (e.g., \(\mathbf{A}, \mathbf{B}, \mathbf{C}, \Delta\)) are stored in the GPU’s high-bandwidth memory (HBM), but only the necessary data from the current chunk is loaded into faster on-chip memory (SRAM or shared memory) for computation, minimizing memory bandwidth bottlenecks.
\end{itemize}

This approach significantly accelerates the computation by reducing the memory overhead and enabling more parallelism.

\subsubsection{Complexity Comparison.}
\begin{itemize}
    \item \textbf{Time Complexity:} Mamba achieves an ideal time complexity of \(O(N L d)\) (where \(N\) is the hidden state dimension, and \(d\) is the input-output channel size), compared to the \(O(L \log L)\) complexity of convolution-based methods. This makes it more advantageous for very long sequences, especially in GPU-parallel environments.
    \item \textbf{Space Complexity:} Through chunking and recomputation techniques, the memory usage is reduced to \(O(\text{batch size} \times d \times \text{chunk size})\), making it scalable for long sequences.
\end{itemize}

\section{Classical Hawkes Processes: A Detailed Exposition}\label{classical HP}
\label{Hawkes}
\subsection{Preliminaries: Counting and Point Processes}
\label{Hawkes1}
We begin by formally defining the foundational concepts of counting and point processes.

\begin{definition}[Counting Process]
A stochastic process $\{N(t), t \geq 0\}$ is defined as a {\bf counting process} if it satisfies the following conditions:
\begin{itemize}
    \item $N(0) = 0$ almost surely (a.s.).  This signifies that the counting process begins with no observed events.
    \item $N(t) \in \{0, 1, 2, \dots\}$ for all $t \geq 0$.  The process only takes on non-negative integer values, representing the number of events that have occurred.
    \item  If $s < t$, then $N(s) \leq N(t)$ a.s. The process is non-decreasing; the number of events can only increase or remain the same.
    \item $N(t) - N(s)$ represents the number of events that occur in the time interval $(s, t]$.
    \item $N(t)$ is right-continuous with left limits (rcll or cadlag).  This means that $N(t) = \lim_{s \to t^+} N(s)$ and the left limit $\lim_{s \to t^-} N(s)$ exists.  Right-continuity ensures that the counting process includes events at the right endpoint.
    \item The process increases by jumps of size +1. Each event increments the counter by exactly one.
\end{itemize}
\end{definition}

\begin{definition}[Point Process]
A {\bf point process} is a random sequence of points (or events) $\{T_i\}_{i=1}^\infty$ in time, where $0 < T_1 < T_2 < \dots < \infty$ a.s.
\begin{itemize}
    \item $T_i$ denotes the time of the $i$-th event occurrence.
    \item The point process is assumed to be simple, meaning no two events occur at the same time, i.e., $T_i \neq T_j$ for $i \neq j$.
    \item It is common to define $T_0 = 0$ for convenience, even though this event isn't counted by the counting process.
\end{itemize}
\end{definition}

\begin{definition}[Relationship: Counting \& Point Processes]
The counting process $N(t)$ can be constructed from the point process $\{T_i\}_{i=1}^\infty$ as follows:
\begin{equation}
    N(t) = \sum_{i=1}^\infty \mathbb{I}(T_i \leq t) = \sum_{T_i \leq t} 1,
\end{equation}
where $\mathbb{I}(A)$ is the indicator function, equal to 1 if event $A$ occurs and 0 otherwise.  This essentially counts the number of events $T_i$ that have occurred up to time $t$.
\end{definition}

\begin{definition}[Inter-Arrival Times]
The inter-arrival times $\{E_i\}_{i=1}^\infty$ are defined as $E_i = T_i - T_{i-1}$ for $i \geq 1$, with $T_0 = 0$. Inter-arrival times represent the time between consecutive events.
\end{definition}

\begin{example}[Homogeneous Poisson Process]
The simplest example of a counting process is the {homogeneous Poisson process} with rate $\lambda > 0$.  In this case, the inter-arrival times are independent and identically distributed (i.i.d.) exponential random variables: $E_i$ i.i.d. $\text{Exponential}(\lambda)$. This means:
\begin{itemize}
    \item The probability density function (pdf) of each $E_i$ is given by $f_{E_i}(x) = \lambda e^{-\lambda x}$ for $x \geq 0$.
    \item The expected value of each inter-arrival time is $\mathbb{E}[E_i] = \frac{1}{\lambda}$.
    \item The expected number of events per unit time is $\mathbb{E}\left[\frac{N(t)}{t}\right] = \lambda$.  This means that the long-run average rate of event occurrence is $\lambda$.
\end{itemize}
\end{example}

\subsection{Conditional Intensity and Compensator}
\label{Hawkes2}
These concepts provide a fundamental way to characterize point processes and their dependence on their past history.

\begin{definition}[Filtration]
Let $(\Omega, \mathcal{F}, \mathbb{P})$ be a probability space. A {\bf filtration} $\{\mathcal{H}_t\}_{t \geq 0}$ is a family of $\sigma$-algebras such that:
\begin{itemize}
    \item $\mathcal{H}_t \subseteq \mathcal{F}$ for all $t \geq 0$.  Each $\mathcal{H}_t$ contains information about events that are measurable in the original probability space.
    \item $\mathcal{H}_s \subseteq \mathcal{H}_t$ for all $0 \leq s \leq t$.  The information available increases (or stays the same) as time progresses.
\end{itemize}
Intuitively, $\mathcal{H}_t$ represents all the information known about the process up to time $t$.
\end{definition}

\begin{definition}[History of a Counting Process]
The {\bf history} of a counting process $\{N(t), t \geq 0\}$, denoted as $\{\mathcal{H}_t\}_{t \geq 0}$, is the filtration generated by the process:
\begin{equation}
    \mathcal{H}_t = \sigma(\{N(s) : 0 \leq s < t\}).
\end{equation}

This means $\mathcal{H}_t$ contains all information about the path of the process $N(s)$ for all times $s$ strictly less than $t$. Note the strict inequality.
\end{definition}

\begin{definition}[Conditional Intensity Process]
The {\bf conditional intensity process} of a counting process $N(t)$ is defined as
\begin{equation}
    \lambda^*(t) = \lim_{\Delta \to 0} \frac{\mathbb{E}(N_{t+\Delta} - N_t \mid \mathcal{H}_t)}{\Delta},
\end{equation}
if this limit exists. This limit represents the instantaneous rate of event occurrence at time $t$, given the history of the process up to (but not including) time $t$.

\item Alternatively, we can write:
\begin{equation}
         \mathbb{P}(dN(t)=1 \mid \mathcal{H}_t) = \lambda^*(t) dt + o(dt),
\end{equation}
where $dN(t) = N(t+dt)-N(t)$ and $o(dt)$ is a term that goes to zero faster than $dt$ as $dt \to 0$.
\end{definition}

\begin{remark}  { }
IIt is crucial to note that $\lambda^*(t)$ is generally a stochastic process itself, reflecting the fact that the rate of event occurrence can change randomly over time based on the history. The superscript * is used to emphasize its random nature.  The conditional intensity is typically left-continuous with right limits (lcrl or caglad).
\end{remark}

\begin{example}[Poisson Process and Conditional Intensity]
If the conditional intensity process $\lambda^*(t) \equiv \lambda(t)$ is a deterministic function of time, meaning it doesn't depend on the history $\mathcal{H}_t$, then the counting process $N(t)$ is a {\bf Poisson process} with rate function $\lambda(t)$.  If $\lambda(t) \equiv \lambda$ is a constant (independent of time), then it is a {\bf homogeneous Poisson process}.
\end{example}

\begin{definition}[Compensator]
The {\bf compensator} of a point process $N(t)$ is defined as
\begin{equation}
   \Lambda_t = \int_0^t \lambda^*(s) \, ds. 
\end{equation}

The compensator represents the expected cumulative number of events up to time $t$, given the history of the process.
\end{definition}

\begin{theorem}[Martingale Characterization of the Compensator]
The compensator $\Lambda_t$ uniquely characterizes the point process $N(t)$.  Specifically, $\Lambda_t$ is the unique predictable process for which the difference
\begin{equation}
   M_t = N_t - \Lambda_t 
\end{equation}

\noindent is a local martingale with respect to the filtration $\mathcal{H}_t$.  This means that
\begin{itemize}
    \item $\mathbb{E}[|M_t|] < \infty$ for all $t \geq 0$.
    \item $\mathbb{E}[M_t \mid \mathcal{H}_s] = M_s$ for all $0 \leq s \leq t$.
\end{itemize}

\item $M_t$ represents a ``fair game", where the expected future value, given past information, is equal to the current value. The compensator removes the predictable part of the process, leaving only the random, unpredictable component.
\end{theorem}

\begin{remark}{ }
IIf $\Lambda_t$ is absolutely continuous with respect to $t$, its pathwise derivative is $\lambda^*(t)$, recovering the conditional intensity interpretation. Moreover, $N(t)$ is a Poisson process if and only if $\Lambda_t$ is a deterministic function of $t$.
\end{remark}

\subsection{The Hawkes Process: Definition and Self-Excitation}
\label{Hawkes3}
We now define the Hawkes process, highlighting its key property of self-excitation.

\begin{definition}[Hawkes Process]
A {\bf Hawkes process} is a counting process $\{N(t), t \geq 0\}$ whose conditional intensity process is given by
\begin{equation}
    \lambda^*(t) = \lambda + \sum_{T_i < t} \mu(t - T_i),
\end{equation}
where:
\begin{itemize}
    \item $\lambda > 0$ is the {\bf background arrival rate}. This represents the baseline rate of events occurring independently of past events.
    \item $\mu: \mathbb{R}_+ \to \mathbb{R}_+$ is the {\bf excitation function} or {\bf kernel}. This function determines how past events influence the current intensity.
\end{itemize}
\end{definition}

The excitation function $\mu(t)$ is the cornerstone of self-excitation. It quantifies the increase in the intensity process at time $t$ due to an event that occurred at time $T_i < t$.

\begin{definition}[Self-Excitation]
The property of {\bf self-excitation} in a Hawkes process arises because the occurrence of an event at time $T_i$ instantaneously increases the conditional intensity $\lambda^*(t)$ for $t > T_i$ by an amount determined by the excitation function $\mu(t - T_i)$. This increase in intensity makes subsequent events more likely to occur in the near future, leading to clustering of events.
\end{definition}

\begin{example}[Exponentially Decaying Hawkes Process]
A common and analytically tractable choice for the excitation function is the exponentially decaying kernel:
\begin{equation}
    \mu(t) = \alpha e^{-\beta t},
\end{equation}
where $\alpha > 0$ controls the magnitude of the excitation, and $\beta > 0$ controls the rate of decay of the excitation. This means that the influence of past events diminishes exponentially over time.
\end{example}

\subsection{Non-homogeneous Poisson Process and Thinning Algorithm}
Non-homogeneous Poisson Process and Ogata’s Thinning Algorithm as follows:
\begin{algorithm}[H]
\caption{Ogata's Thinning Algorithm for Non-homogeneous Poisson Process}
\label{alg:ogata_thinning}
\textbf{Input:} Intensity function \(\lambda_k(t)\), start time \(T_{\text{orig}}\), end time \(T_{\text{end}}\) \\
\textbf{Output:} Padded event sequence \(\mathcal{T}_{\text{pad}}^{(k)} = \{t_{\text{pad}}^{(1)}, \cdots, t_{\text{pad}}^{(N_k)}\}\)
\begin{algorithmic}[1]
\State \textbf{Step 1: Determine an Upper Bound}
\State Find a constant \(\lambda_{\max}^{(k)}\) such that \(\lambda_k(t) \le \lambda_{\max}^{(k)} \quad \text{for all } t \in [T_{\text{orig}}, T_{\text{end}}]\)
\State \textbf{Step 2: Generate Candidate Events}
\State Generate inter-arrival times \(\Delta t \sim \mathrm{Exp}\bigl(\lambda_{\max}^{(k)}\bigr)\)
\State Initialize \(t_{\text{cand}} \gets T_{\text{orig}}\)
\State Initialize \(i \gets 1\)
\While{$t_{\text{cand}} < T_{\text{end}}$}
    \State $\Delta t_i \sim \mathrm{Exp}\bigl(\lambda_{\max}^{(k)}\bigr)$
    \State $t_{\text{cand}}^{(i)} \gets t_{\text{cand}} + \Delta t_i$
    \State \textbf{Step 3: Accept/Reject Candidate Events}
    \State Acceptance probability \(p_{\text{accept}}^{(k)}\Bigl(t_{\text{cand}}^{(i)}\Bigr) = \frac{\lambda_k\Bigl(t_{\text{cand}}^{(i)}\Bigr)}{\lambda_{\max}^{(k)}}\)
    \State Generate a uniform random number \( u \sim \mathcal{U}(0,1) \)
    \If{$u \le p_{\text{accept}}^{(k)}\Bigl(t_{\text{cand}}^{(i)}\Bigr)$}
        \State Accept candidate event: $t_{\text{pad}}^{(i)} \gets t_{\text{cand}}^{(i)}$
        \State $i \gets i + 1$
    \EndIf
    \State $t_{\text{cand}} \gets t_{\text{cand}}^{(i)}$
\EndWhile
\State \textbf{Step 4: Construct the Padding Event Sequence}
\State \(\mathcal{T}_{\text{pad}}^{(k)} = \{t_{\text{pad}}^{(1)}, t_{\text{pad}}^{(2)}, \ldots, t_{\text{pad}}^{(N_k)}\}\), where \(N_k\) is the number of accepted events
\If{number of generated events exceeds $T_{\text{pad}} = T_{\text{target}} - T_{\text{orig}}$}
    \State Use only the first \(T_{\text{pad}}\) events
\EndIf
\State \Return \(\mathcal{T}_{\text{pad}}^{(k)}\)
\end{algorithmic}
\end{algorithm}

\subsection{The Immigration-Birth View and Stationarity}
\label{Hawkes4}
The immigration-birth representation provides an intuitive way to understand the Hawkes process as a Poisson cluster process.

\begin{itemize}
    \item {\bf Immigration Process:} A homogeneous Poisson process with rate $\lambda$. Events in this process represent exogenous shocks or external events that trigger the Hawkes process.
    \item {\bf Birth Process:} Each immigration event occurring at time $s$ generates a new inhomogeneous Poisson process (also known as a Poisson point process) with intensity $\mu(t - s)$ for $t > s$. These events represent offspring or triggered events caused by the initial immigration event.
    \item {\bf Hawkes Process:} The Hawkes process is the superposition of the immigration process and all the birth processes.
\end{itemize}

This representation allows us to analyze the Hawkes process using tools from branching process theory.

\begin{definition}[Branching Ratio]
An event occurring at time $s$ will generate a Poisson-distributed number of first-generation offspring over the time interval $[s, \infty)$. The expected number of offspring, known as the branching ratio, is defined as:
\begin{equation}
    \eta = \int_{s}^\infty \mu(t - s) \, dt  = \int_0^\infty \mu(t) \, dt.
\end{equation}

\end{definition}

\begin{theorem}[Stationarity Condition]
The Hawkes process is stationary (i.e., the process does not explode) and has a finite mean if and only if the branching ratio is less than one:
\begin{equation}
    \eta < 1
\end{equation}

This condition ensures that, on average, each event generates less than one offspring, preventing the process from growing unbounded.
\end{theorem}

\begin{example}[Stationarity]
If $\mu(t) = \alpha e^{-\beta t}$, then the branching ratio is:
\begin{equation}
    \eta = \int_0^\infty \alpha e^{-\beta t} \, dt = \frac{\alpha}{\beta}.
\end{equation}

Therefore, the process is stationary if and only if $\alpha < \beta$.
\end{example}

\begin{theorem}[Law of Large Numbers]
Under the stationarity condition ($\eta < 1$), the long-term average rate of events converges to:
\begin{equation}
    \lim_{t \to \infty} \frac{N_t}{t} = \frac{\lambda}{1 - \eta} \quad \text{a.s.}
\end{equation}

This result provides an explicit formula for the long-run average rate in terms of the background rate and branching ratio.
\end{theorem}

\subsection{Excitation Functions and Markov Analysis}
\label{Hawkes5}
The choice of excitation function $\mu(t)$ is critical for modeling specific characteristics of the self-exciting process.

\begin{itemize}
    \item Delayed Impact: The excitation function can be designed to incorporate a delayed impact, where the influence of an event is not immediate but increases over time. This can be achieved using functions such as the gamma distribution pdf for $\mu(t)$.
    \item Domain-Specific Expertise:  The excitation function can be tailored to incorporate domain knowledge. The Omori-Utsu law in seismology is an example of a power-law kernel used to model aftershock rates.
\end{itemize}

\begin{definition}[Omori-Utsu Law]
The Omori-Utsu law models the aftershock rate following an earthquake:
\begin{equation}
    \mu(t) = \frac{K}{(t + c)^p},
\end{equation}
where $K, c > 0$ and $p > 1$. The parameters control the magnitude, shift, and decay rate of the aftershock sequence.
\end{definition}

The exponentially decaying excitation function $\mu(t) = \alpha e^{-\beta t}$ is particularly important due to the Markov property.

\begin{theorem}[Markov Property]
When the excitation function is exponentially decaying, the joint process $(N_t, \lambda^*_t)$ is a Markov process. This simplifies inference, simulation, and analysis because the future state of the process depends only on the current state and not on the entire past history.

\item In particular, the conditional intensity between jumps ($T_n \leq t < T_{n+1}$) follows:
\begin{equation}
    \lambda^*_t = \lambda + (\lambda^*_{T_n+} - \lambda) e^{-\beta (t - T_n)} = \lambda + (\lambda^*_{T_n} + \alpha - \lambda) e^{-\beta (t - T_n)}.
\end{equation}

\end{theorem}

\begin{definition}[Exponentially Decaying Hawkes Process]
An exponentially decaying Hawkes process can be further generalized by allowing the conditional intensity to start at an initial value $\lambda_0$ that differs from the background rate $\lambda$.  The conditional intensity for $t \geq 0$ is given by
\begin{equation}
    \lambda^*_t = \lambda + (\lambda_0 - \lambda) e^{-\beta t} + \sum_{T_i < t} \alpha e^{-\beta (t - T_i)}.
\end{equation}

The compensator for this process is:
\begin{equation}
    \Lambda_t = \lambda t + \frac{(\lambda_0 - \lambda)}{\beta} (1 - e^{-\beta t}) + \sum_{T_i < t} \frac{\alpha}{\beta} (1 - e^{-\beta (t - T_i)}).
\end{equation}

\end{definition}

\subsection{Likelihood-Based Inference}
\label{Hawkes6}
Maximum likelihood estimation (MLE) is a standard method for parameter estimation in Hawkes processes.

\begin{theorem}[Likelihood Function for Point Processes]
The likelihood function for a point process with parameter vector $\theta$, given observations $\{t_1, \dots, t_n\}$ within a time horizon $[0, T]$, is:

\begin{align}
L(\theta \mid t_1, \dots, t_n, T) &= \left[ \prod_{i=1}^n \lambda^*_{t_i} \right] \exp\left(-\int_0^T \lambda^*_t \, dt\right) \notag \\ 
&= \left[ \prod_{i=1}^n \lambda^*_{t_i} \right] e^{-\Lambda_T}.
\end{align}
\end{theorem}

\begin{theorem}[Log-Likelihood Function]
The corresponding log-likelihood function is:
\begin{equation}
    \ell(\theta \mid t_1, \dots, t_n, T) = \sum_{i=1}^n \log \lambda^*_{t_i} - \Lambda_T.
\end{equation}

\end{theorem}

\begin{theorem}[Log-Likelihood for Hawkes Process]
Substituting the Hawkes process conditional intensity, the log-likelihood function becomes:
\begin{equation}
    \ell(\theta \mid t_1, \dots, t_n, T) = \sum_{i=1}^n \log \left( \lambda + \sum_{t_j < t_i} \mu(t_i - t_j) \right) - \Lambda_T.
\end{equation}
\end{theorem}

Direct numerical maximization of the log-likelihood is a common approach for parameter estimation. However, it may encounter convergence issues if the log-likelihood function is flat near its maximum.

\subsection{In-Depth Discussion of the Intensity Function}
\label{Intensity Function}

The intensity function defined in Equation (\ref{eq:mhp_intensity}) is a sophisticated model designed to capture the complex dynamics of event sequences in micro-videos, particularly the cross-influence between different types of events. Let's dissect the function to understand its key components and motivations:

\subsubsection{Core Motivation: Modeling Interdependent Event Sequences.}

The primary motivation behind this particular formulation stems from the observation that events in social media, such as likes, shares, comments, and favorites, are not independent. The occurrence of one type of event can significantly influence the probability of subsequent events of the same type (self-excitation) or different types (cross-excitation). Capturing these interdependencies is crucial for accurate modeling and prediction.

\subsubsection{Deconstructing the Components.}

Let's examine each term of the intensity function in detail:

\begin{itemize}
    \item \textbf{Baseline Intensity (\(\mu_k\)):}  The term \(\mu_k\) represents the intrinsic or baseline rate at which events of type \(k\) occur, independent of any past events. It accounts for factors such as the inherent appeal of the video or general user activity. This is a constant term, and guarantees that the conditional intensity is always positive.

    \item \textbf{Cross-Excitation Term \\(\(\sum_{m=1}^{4} \sum_{h: t_h^{m} < t} \alpha_{k,m} \exp\Bigl(-\delta_{k,m}(t - t_h^{m})\Bigr)\)):} This is the heart of the model's ability to represent cross-excitation.

        \begin{itemize}
            \item \textbf{Inner Sum (\(\sum_{h: t_h^{m} < t}\)):} This sum iterates through all historical events of type \(m\) that occurred before time \(t\). Each historical event contributes to the current intensity of event type \(k\).

            \item \textbf{Cross-Excitation Coefficient (\(\alpha_{k,m}\)):} The coefficient \(\alpha_{k,m}\) is crucial. It quantifies the degree to which an event of type \(m\) influences the occurrence of events of type \(k\).
                \begin{itemize}
                    \item If \(\alpha_{k,m} > 0\), it indicates that an event of type \(m\) \textit{excites} (promotes or increases the likelihood of) events of type \(k\). For example, a high number of likes (\(m\) = likes) might encourage more shares (\(k\) = shares).
                    \item If \(\alpha_{k,m} < 0\), it suggests that an event of type \(m\) \textit{inhibits} (decreases the likelihood of) events of type \(k\). This could happen in some scenarios; e.g., a large number of comments (which may include negative feedback) might discourage further likes. Although, in this function, it's assumed that \(\alpha_{k,m} > 0\).
                    \item If \(\alpha_{k,m} \approx 0\), it means that events of type \(m\) have little to no direct influence on events of type \(k\).
                    \item The magnitude of \(\alpha_{k,m}\) reflects the strength of the influence. Larger absolute values imply stronger effects.
                \end{itemize}

            \item \textbf{Exponential Decay Function \\(\(\exp\Bigl(-\delta_{k,m}(t - t_h^{m})\Bigr)\)):} This function models the decaying influence of past events over time.  The parameter \(\delta_{k,m}\) determines the rate of decay.
                \begin{itemize}
                    \item A larger \(\delta_{k,m}\) implies a faster decay rate. The influence of the past event quickly diminishes.
                    \item A smaller \(\delta_{k,m}\) indicates a slower decay rate. The influence of the past event persists for a longer duration.
                    \item The exponential form is chosen because it's a simple, mathematically tractable way to represent this decaying influence. It captures the idea that recent events have a stronger impact than distant events.
                \end{itemize}
        \end{itemize}

    \item \textbf{Neural Network Term (\(f_k\bigl(h(t)\bigr)\)):} This term introduces a non-linear component, capturing more complex dependencies and hidden factors that cannot be adequately modeled by the linear superposition of past event influences. This is typically used to inject features of the video into the conditional intensity calculation. The  \(h(t)\) represents some hidden state or representation learned by a neural network (Mamba in this specific instance) up to time \(t\).
\end{itemize}

\subsubsection{Advantages of this Formulation.}

\begin{itemize}
    \item \textbf{Flexibility:} This intensity function offers considerable flexibility in capturing a wide range of dependencies between different types of events.
    \item \textbf{Interpretability:} The parameters (\(\mu_k\), \(\alpha_{k,m}\), \(\delta_{k,m}\)) have clear interpretations, providing insights into the underlying dynamics of the event sequences.
    \item \textbf{Adaptability:} The integration of the neural network term allows the model to learn and adapt to complex, nonlinear patterns in the data.
\end{itemize}

\subsubsection{Key Assumption: Additivity.}

A core assumption of this formulation is that the influences of past events are additive (through the summation). While this simplifies the model, it might not always be perfectly realistic. In some scenarios, event influences might interact in more complex, non-additive ways.

\subsubsection{In Summary: A Holistic Model.}

This intensity function presents a holistic approach to modeling event sequences by: (1) establishing a basic arrival rate, (2) capturing the effect of past events from various sources that influence the present behavior, and (3) Incorporating intricate, high-order dependencies that may be discovered directly from data. Overall, it integrates interpretability with a more nuanced nonlinear component, providing a practical approach for event modeling.

\subsection{Derivation and Rationale for the MHP Intensity Function}
\label{sec:appendix_mhp_derivation}

The formulation of the Mamba-Hawkes Process (MHP) intensity function, presented in Equation~(1), represents a principled synthesis of the classical Hawkes process (Appendix~\ref{classical HP}) and the Mamba state-space model (Appendix~\ref{app:architecture}). This section provides a detailed derivation to elucidate the connection between these components, justifying our model's architecture.

A classical multivariate Hawkes process, as detailed in Appendix C.3, models the event intensity as a linear superposition of a baseline rate and contributions from past events, typically via an exponential decay kernel. This formulation, while interpretable, is predicated on the strong assumption that the influence of historical events is both additive and context-agnostic. That is, the impact of an event is predetermined by its type and time of occurrence, irrespective of the broader sequence of interactions in which it is embedded. This assumption is often violated in real-world scenarios like micro-video engagement, where the dynamics are highly non-linear. For instance, an initial surge of likes might be positively reinforcing, but their influence could be completely negated or even reversed by a subsequent wave of negative comments. A classical Hawkes model is structurally incapable of capturing such complex, context-dependent modulations of influence.

To address this fundamental limitation, we augment the classical Hawkes formulation with a dynamic, non-linear component derived from a Mamba model. As described in Appendix B, Mamba is a selective state-space model adept at capturing long-range dependencies in sequential data. When applied to a sequence of user interactions $\mathcal{R} = \{(t_0, y_0), \dots, (t_n, y_n)\}$, the Mamba module processes the entire event history and distills it into a compressed, history-aware hidden state vector, $h(t)$. This vector is not a mere aggregation but a rich representation that encodes the intricate sequential and causal dependencies among past events.

The bridge between the two paradigms is the function $f_k(h(t))$ in our MHP intensity function. We posit that the full intensity of an event is determined not only by the linear excitation from past events but also by a non-linear correction based on the global context of the event history. The Mamba-generated state $h(t)$ provides this global context. The term $f_k(h(t))$, which is implemented as a small neural network that takes $h(t)$ as input, serves as this dynamic correction factor. It can output positive or negative values, allowing it to adaptively increase or decrease the event intensity. For example, if $h(t)$ represents a history of waning engagement, $f_k(h(t))$ can output a negative value that suppresses the intensity calculated by the classical Hawkes components, thereby accurately modeling a ``Later Decline" phase.

Thus, the complete MHP intensity function from Equation~(1),
\begin{equation}
    \lambda_k(t) = \mu_k + \sum_{m=1}^{M}\sum_{t_i^{m} < t} \alpha_{k,m} e^{-\delta_{k,m}(t-t_i^{m})} + f_k(h(t)), \nonumber
\end{equation}
is a composite model. It retains the interpretable baseline intensity ($\mu_k$) and the linear, exponentially decaying influence from past events (the summation term) of the classical Hawkes process. Crucially, it enhances this foundation with the term $f_k(h(t))$, which leverages the powerful sequence modeling capabilities of Mamba to introduce a non-linear, context-aware correction. This synergistic integration allows the MHP to capture the nuanced, dynamic, and often non-monotonic trajectories of micro-video popularity far more faithfully than either of its constituent models could alone.

\section{Justification for Using Ogata's Thinning Algorithm for Temporal Embeddings}
\label{app:thinning_justification}
In this section, we provide a detailed justification for our choice to employ Ogata's thinning algorithm to generate fixed-length temporal embeddings from the Mamba-Hawkes Process (MHP). This design choice was rightly questioned, noting that modern architectures like Mamba can inherently process variable-length sequences. Our rationale is rooted in the specific requirements of the downstream \textbf{temporal-aware retrieval module} and our objective to maintain the highest possible fidelity in the temporal representation while enabling efficient, large-scale similarity search.

\subsection{The Core Challenge: Efficient and Meaningful Similarity Search}
The primary goal of our retrieval module is to identify historical videos that are similar to a target video along two axes: multi-modal content and \textit{temporal popularity dynamics}. For a retrieval system to be practical, especially when querying a large memory bank ($B$) containing millions of candidates, the similarity computation must be highly efficient. The dot product, as used in our temporal similarity calculation `SStemp` (Equation 3), is one of the most efficient methods for comparing two vector representations. It forms the backbone of modern large-scale vector search systems (e.g., FAISS, ScaNN). This operation, however, fundamentally requires that the two vectors being compared have a fixed, identical dimensionality.

This requirement presents a direct conflict: our Mamba-Hawkes Process (MHP) naturally produces a variable-length sequence of hidden states corresponding to the variable number of user interaction events, but our retrieval mechanism is most efficient with fixed-length vectors. The crucial question then becomes: how do we best transform a variable-length, event-driven temporal representation into a fixed-length vector with minimal loss of information?

\subsection{Limitations of Naive Alternatives (e.g., Pooling)}
A common approach to creating a fixed-size representation from a variable-length sequence is to use a pooling operation (e.g., mean-pooling or max-pooling) on the sequence of MHP outputs. We argue that such an approach, while simple, is fundamentally inadequate for our task and would undermine the very motivation for using a sophisticated temporal model like the MHP.

\begin{itemize}
    \item \textbf{Mean-Pooling:} Applying mean-pooling would average the temporal dynamics over the entire event sequence. This would effectively smooth out and destroy the precise ``rollercoaster-like" patterns—the initial surges, the subsequent decays, the bursts of activity—that we aim to capture. A video with a sharp early peak and rapid decline would become indistinguishable from a video with slow, steady growth if their average event intensity is similar. This is a highly ``lossy" compression that discards the most discriminative temporal information.
    
    \item \textbf{Max-Pooling:} Max-pooling would only capture the single most intense moment of user interaction, ignoring the overall shape, duration, and decay characteristics of the popularity curve. It cannot differentiate between a flash-in-the-pan viral event and one that builds to a peak and sustains interest.
    
    \item \textbf{Last Hidden State:} Using the last hidden state of the Mamba model is another common technique. While this encodes information from the entire sequence, its representation is heavily biased towards the most recent events, failing to provide a balanced summary of the overall temporal trajectory, especially the crucial early-stage dynamics.
\end{itemize}

In short, while these simpler alternatives would solve the fixed-length problem, they would do so by discarding the nuanced, sequential, and causal information that our MHP module is specifically designed to model.

\subsection{Ogata's Thinning as a Principled, High-Fidelity Solution}
In contrast to the naive methods above, Ogata's thinning algorithm ~\cite{thinning} is not merely a ``padding" technique. It is a \textbf{statistically principled simulation method}. Instead of adding meaningless zero-vectors or applying a lossy pooling function, the thinning algorithm generates a new, longer event sequence that is a \textit{statistically faithful realization} of the very same Non-Homogeneous Poisson Process (NHPP) learned by our MHP model.

The process works as follows:
\begin{itemize}
    \item We first learn the complex, time-varying intensity function $\lambda_k(t)$ for a given video using our MHP module. This function represents the instantaneous probability of an event of type $k$ occurring at any time $t$.
    
    \item We establish a constant upper bound, $\lambda_{\text{max}}$, for this intensity function over the time interval of interest.
    
    \item We then generate candidate events from a simpler, homogeneous Poisson process with a constant rate $\lambda_{\text{max}}$.
    
    \item Each candidate event at time $t_{\text{cand}}$ is ``thinned" (i.e., accepted or rejected) with a probability of $\lambda_k(t_{\text{cand}}) / \lambda_{\text{max}}$.
\end{itemize}

The resulting sequence of accepted events is guaranteed to be a valid sample from the original, complex NHPP defined by $\lambda_k(t)$. By running this simulation until we obtain a pre-defined number of events (our fixed length), we are effectively ``completing" the popularity trajectory according to the learned dynamics.

\subsection{Maintaining Fidelity and Creating a Canonical Representation}
The key advantage of this approach is its \textbf{high fidelity}. The simulated events are not noise; they are plausible future events that \textit{could have happened} according to the model's understanding of the video's temporal dynamics. This allows us to create what we term a \textbf{canonical temporal signature}. By simulating all interaction sequences to the same fixed length, we are standardizing them not based on their observed history (which can be sparse or short), but based on the underlying temporal process they represent.

This provides a much richer and more robust basis for comparison:
\begin{itemize}
    \item A video with a high initial intensity will be simulated to have more events early in its canonical signature.
    \item A video with a slow decay rate will show a more sustained pattern of events throughout its signature.
\end{itemize}
This allows our dot-product similarity to directly compare the fundamental temporal \textit{shapes} of popularity curves, which was our original goal.

\subsection{Conclusion}
In summary, the decision to use Ogata's thinning algorithm was a deliberate engineering and modeling choice to resolve the tension between the need for a high-fidelity, variable-length temporal representation and the practical requirement for an efficient, fixed-length vector for large-scale retrieval. Simpler methods like pooling were considered and rejected due to their lossy nature, which would compromise the core contribution of our temporal modeling. The thinning algorithm provides a statistically sound method to generate a fixed-length canonical temporal signature, preserving the intricate dynamics captured by the MHP and enabling meaningful and efficient similarity calculations. We thank the reviewer for this insightful question, which has allowed us to clarify this critical aspect of our methodology. This careful balance between representation fidelity and retrieval efficiency is a key contribution of our work.

\section{Multi-Modal Feature Extraction}\label{app:extraction}
For MVPP, it is fatal to extract the required feature information from the multi-modal information of the video. We perform frame-by-frame processing on all target videos, select certain interval frames for extracting image and audio information, and transform them into corresponding frame feature vectors through Vision Transformer 
(ViT)~\cite{VIT} and Audio Spectrogram Transformer (AST) architectures~\cite{ast}. Further, we input the image-audio features of the video and the text description into a model with both forward and reverse attention mechanisms to capture its content features. Further, we compare it with the videos matched by the retriever in the database to further enhance this feature. \par

To capture the key features of a micro video $V_i$, we extract visual, audio and textual information. We start by selecting key frames $F_1^i, F_2^i, \dots, F_K^i$, where $K$ is the total number of frames. Each frame $F_j^i \in \mathbb{R}^{H\times W\times C}$ is divided into fixed-size patches and reshaped ${F_j^i}^\ast \in \mathbb{R}^{N_p\times P^2\times C}$, where $(H, W)$ is the resolution of the original frame, $C$ is the number of channels, $(P, P)$ is the patch size, and $N_p$ is the number of patches, which serves as the input sequence length for the Vision Transformer (ViT)~\cite{VIT}. Through ViT, visual information of video frames is encoded into visual embedding $E_i^v\in \mathbb{R}^{K \times d_v}$, where $d_a$ denotes the length of a single embedding.\par
For audio features, we follow a similar approach using the Audio Spectrogram Transformer (AST). We first align the timestamps of the sampled frames with corresponding audio segments $W = \{w_1, w_2, \dots, w_K\}$ extracted from the video’s soundtrack. These audio segments are processed using a Fast Fourier Transform (FFT)~\cite{fast} to generate Mel-frequency spectral coefficients (MFCC), resulting in a spectrum matrix $M_S\in \mathbb{R}^{n_{mel} \times n_a}$, where $n_{mel}$ is the number of mel-filters, and $n_a$ is the number of time frames, determined by the audio length and stride. This spectrum matrix, treated as a single-channel image, is then processed using the same patching method applied to visual frames and transformed into acoustic embedding $E_i^a \in \mathbb{R}^{K \times d_a}$, where $d_a$ denotes the length of a single embedding.\par
Next, these feature matrices are concatenated to emphasize the relationship between visual and audio information:
\begin{equation}
        E_i^{v,a} =E_i^v\oplus E_i^a\in \mathbb{R}^{K\times (d_v+d_a)}.
\end{equation}

The concatenated features $E_i^{v,a}$ are then passed through a linear layer $W_c \in \mathbb{R}^{(d_v+d_a)\times d}$ with a ReLU activation function, generating an audio-visual embedding $X^c_i$. Similarly, we can get $E^t_i \in \mathbb{R}^{n_w \times d_t}$ through a pre-trained sentence embedding model AnglE, where $n_w$ is the word length in the textual descriptions and $d_t$ denotes the length of an embedded token. And then we acquire the textual embedding $X_i^t$ through a linear layer $W_t \in \mathbb{R}^{d_t\times d}$:
\begin{align}
    X_i^c=ReLU(E_i^{v,a} W_c)\in \mathbb{R}^{K\times d},\\ 
    X_i^{t}=ReLU(E_i^{t}W_t)\in \mathbb{R}^{n_w\times d}.
\end{align}

\section{Feature Fusion and Popularity Prediction}
\label{app:interaction_and_prediction}
\subsection{Cross-modal/temporal Attention}
Aligning audio-visual, textual modalities and temporal series of micro videos presents challenges due to potential inconsistencies between these three kinds of features: textual descriptions, audio-visual content and user feedback time-series. To address this, we implement three pairs of cross-attention network across three modal/temporal. The cross modal-temporal attention comprises both positive and negative attention modules, designed to capture the similarities and differences between multi-modal/temporal information. The positive attention module focuses on identifying the most consistent features across different modalities, while the negative attention module highlights any inconsistent or contradictory information.\par
To illustrate, we initially examine the cross-attention mechanism between the audio-visual and textual modalities as a representative case. \par
Within the positive attention module, the most aligned features between modalities are calculated using cross-modal/temporal attention vectors. For a given video $V_i$, the audio-visually guided positive textual features $T_i^\mathcal{P}$ and the textually guided positive audio-visual features $C_i^\mathcal{P}$ are derived as follows:
\begin{align}
T_i^\mathcal{P}&=ATT^{\mathcal{P}}(X_i^{c}W^{\mathcal{Q}}_{\mathcal{P}},X_i^{t}W^{\mathcal{K}}_{\mathcal{P}},X_i^{t}W^{\mathcal{V}}_{\mathcal{P}})\notag\\
&=Softmax(\alpha\frac{\mathcal{QK}^T}{\sqrt{d}})\mathcal{V},\\
C_i^\mathcal{P}&=ATT^{\mathcal{P}}(X_i^{t}W^{\mathcal{Q}}_{\mathcal{P}},X_i^{c}W^{\mathcal{K}}_{\mathcal{P}},X_i^{c}W^{\mathcal{V}}_{\mathcal{P}})\notag \\
&=Softmax(\alpha\frac{\mathcal{QK}^T}{\sqrt{d}})\mathcal{V},
\end{align}
where $ATT$ function represents the Attention mechanism, $\alpha$ is the positive scaling factor, $X^c_i$ denotes the embedding generated from audio-visual extraction and $X^t_i$ is the input of textual embedding. Besides, $W^{\mathcal{Q}}_{\mathcal{P}},W^{\mathcal{K}}_{\mathcal{P}},W^{\mathcal{V}}_{\mathcal{P}}$
denote the query, key, and value projection matrices, respectively. The parameter $\alpha$ is used to balance the influence of positive and negative attention. Similarly, the negative audio-visually guided textual features $T_i^\mathcal{N}$ and textual guided audio-visual features $C_i^\mathcal{N}$ can be obtained as follows, where $\gamma=-(1-\alpha)$ is the negative scaling factor:
\begin{align}
T_i^\mathcal{N}&=ATT^{\mathcal{N}}(X_i^{c}W^{\mathcal{Q}}_{\mathcal{N}},X_i^{t}W^{\mathcal{K}}_{\mathcal{N}},X_i^{t}W^{\mathcal{C}}_{\mathcal{N}})\notag \\
&=Softmax(\gamma\frac{\mathcal{QK}^T}{\sqrt{d}})\mathcal{V},\\
C_i^\mathcal{N}&=ATT^{\mathcal{N}}(X_i^{t}W^{\mathcal{Q}}_{\mathcal{N}},X_i^{c}W^{\mathcal{K}}_{\mathcal{N}},X_i^{c}W^{\mathcal{C}}_{\mathcal{N}})\notag \\ &=Softmax(\gamma\frac{\mathcal{QK}^T}{\sqrt{d}})\mathcal{V}.
\end{align}

After this, we referred to the integration processing of hidden states in the MMRA ~\cite{FNN} ~\cite{MMRA} model to generate a comprehensive textual modal representation in FFN layers, incorporate audio-visual hidden states into textual hidden states to generate a comprehensive textual, audio-visual modal representation $\widetilde{T_i}$ and $\widetilde{C_i}$, which modify the calculation of the FFN process as follows:
\begin{align}
    \widetilde{T_i^c}&=ReLU\left( X_i^t+\left( C_i^{\mathcal{P}}\oplus C_i^{\mathcal{N} } \right)W^t_1 \right)W^t_2,\\
    \widetilde{C_i^t}&=ReLU\left( X_i^c+\left(T_i^{\mathcal{P}}\oplus T_i^{\mathcal{N} } \right)W^c_1 \right)W^c_2,
\end{align}
where $W^t_1,W^c_1 \in \mathbb{R}^{2d\times d}$, $W^t_2,W^c_2 \in \mathbb{R}^{d\times d}$ represent learnable weights matrix via the attentive pooling strategy, $\oplus$ denotes the concatenation operation.\par

Similarly, $T_i^s$,$S_i^t$ and $C_i^s$,$S_i^t$ can be obtained through textual-temporal and audiovisual-temporal cross-attention respectively. Finally, we exploit expressive representations $T_i,C_i,S_i$ by concatenation and transformation, followed by the attentive pooling strategy~\cite{APS}. \par

\subsection{Retrieval Enhanced Interaction and Prediction}
We focus on extracting valuable insights from relevant videos retrieved to improve micro video popularity prediction (MVPP). The retrieved videos hold the similar multi-modal/temporal patterns with the target video, which enhance the characteristic of the target video. \par
The retrieved feature from Multi-modal/temporal Retrieval (see section 4.3) can be integrated into the cross-attention structure and effectively learned. We employ the same process on the retrieved embeddings $X_i^{r,c}$, $X_i^{r,t}$ and $X_i^{r,t}$, to calculate $T_i^r$, $C_i^r$ and $S_i^r$. For information augmentation, we also encode the retrieved label embedding $L_i^r$ of each video $V_i$. Finally, we integrate all features from the micro-video $V_i$: $T_i$, $C_i$, $S_i$ and $T_i^r$, $C_i^r$, $S_i^r$, along with the retrieved labels $L_i^r$. Thus, we could construct cross-sample interactions as follows:
$$
    \mathcal{I}=[\Phi(C_i, C_i^r), \Phi(C_i, T_i^r), \Phi(C_i, S_i^r), ..., \Phi(S_i, S_i^r),\Phi(S_i,L_i^r)],
$$
where $\Phi$ denotes the process of inner products. \par
In the end, we put all the components into a prediction network. For the micro-video $V_i$, the output layer is fed with the concatenated vector as $H=concat([T_i, C_i, S_i, T_i^r, C_i^r, S_i^r, L_i^r,  \mathcal{I}])$. The output layer is multilayer perceptrons (MLPs) with one final output unit to predict the popularity(heart, share, play, comment) of the target video. 

\section{Assumptions and Proof Sketch for Theorem 1}
\label{sec:assumptions_proof}

The oracle properties of the MHP estimator, as stated in Theorem~\ref{sec:theorem1}, hold under the following standard assumptions.

\begin{assumption}[Stationarity and Ergodicity]
\label{assump:stationarity}
The true Hawkes process $N(t; \theta^*)$ is stationary and ergodic. This requires the spectral radius of its branching ratio matrix to be less than 1. This ensures that the Law of Large Numbers and Central Limit Theorem apply to the log-likelihood and its derivatives.
\end{assumption}

\begin{assumption}[Model Identifiability]
The parameterization is identifiable. If $\theta_1 \neq \theta_2$, then the corresponding probability laws of the Hawkes processes differ.
\end{assumption}

\begin{assumption}[Regularity Conditions]
\label{assump:regularity}
The log - likelihood function $\ell(\theta)$ is twice continuously differentiable with respect to $\theta$ in a neighborhood of $\theta^*$. The Fisher information matrix $I(\theta) = -E[\nabla^2 \ell(\theta)]$ is finite and positive definite at $\theta = \theta^*$.
\end{assumption}

\begin{assumption}[Penalty Function]
\label{assump:penalty}
The penalty function $p_{\gamma_N}(t)$ is a non-decreasing and concave function for $t > 0$. Its derivative $p'_{\gamma_N}(t)$ must satisfy that for some $a > 1$, $p'_{\gamma_N}(t) = \gamma_N$ for $t \approx 0^+$ and $\lim_{t\to\infty} p'_{\gamma_N}(t) = 0$. The regularization parameter is chosen such that as $N \to \infty$, $\gamma_N \to 0$ and $\sqrt{N}\gamma_N \to \infty$. This is satisfied by the MCP and SCAD penalties.
\end{assumption}
The proof proceeds by analyzing the Karush-Kuhn-Tucker (KKT) conditions for the penalized likelihood estimator \(\hat{\theta}_N\), showing that with probability approaching 1, any solution satisfying these conditions exhibits the desired oracle properties: sparsity recovery and asymptotic normality equivalent to the oracle estimator.

Under Assumptions \ref{assump:stationarity}--\ref{assump:regularity}, it follows from the theory of M-estimators for point processes that there exists a \(\sqrt{N}\)-consistent local maximizer \(\hat{\theta}_N\) of the penalized likelihood function, satisfying \(\|\hat{\theta}_N - \theta^*\| = O_p(N^{-1/2})\). This consistency allows the analysis to be localized to a shrinking neighborhood of the true parameter \(\theta^*\).

A necessary condition for \(\hat{\theta}_N\) to be a local maximizer is that the subgradient of the objective function contains zero, yielding the KKT conditions for \(j = 1, \dots, p\):
\begin{equation}
    \frac{\partial L_N(\hat{\theta}_N)}{\partial \theta_j} - p'_{\gamma_N}(|\hat{\theta}_{N,j}|) \operatorname{sign}(\hat{\theta}_{N,j}) = 0 \quad \text{if } \hat{\theta}_{N,j} \neq 0,
\end{equation}

\begin{equation}
    \left| \frac{\partial L_N(\hat{\theta}_N)}{\partial \theta_j} \right| \le p'_{\gamma_N}(0) \quad \text{if } \hat{\theta}_{N,j} = 0.
\end{equation}

By Assumption \ref{assump:penalty}, \(p'_{\gamma_N}(0) = \gamma_N\). The analysis is divided into the true non-zero coefficients (indexed by \(\mathcal{A}\)) and the true zero coefficients (indexed by \(\mathcal{A}^c\)).

To establish sparsity recovery, we show that \(\hat{\theta}_{N,j} = 0\) for all \(j \in \mathcal{A}^c\) with probability approaching 1. Consider a candidate solution of the form \(\hat{\theta} = (\hat{\theta}_{\mathcal{A}}, \mathbf{0}_{\mathcal{A}^c})\). For this to satisfy the KKT conditions, it must hold that \(\left| \partial L_N(\hat{\theta}) / \partial \theta_j \right| \le \gamma_N\) for each \(j \in \mathcal{A}^c\).

A Taylor expansion of \(\partial L_N(\hat{\theta}) / \partial \theta_j\) around \(\theta^*\) gives
\begin{equation}
   \frac{\partial L_N(\hat{\theta})}{\partial \theta_j} = \frac{\partial L_N(\theta^*)}{\partial \theta_j} + \nabla^2_{\theta_j} L_N(\theta^*)(\hat{\theta} - \theta^*) + o_p(N^{-1/2}). 
\end{equation}

The score term \(\partial L_N(\theta^*) / \partial \theta_j\) is \(O_p(N^{-1/2})\) by the central limit theorem for martingales, while the Hessian term is also \(O_p(N^{-1/2})\) due to the consistency of \(\hat{\theta}\) and the convergence of \(\nabla^2 L_N(\theta^*)\) to the finite Fisher information matrix. Thus, \(\left| \partial L_N(\hat{\theta}) / \partial \theta_j \right| = O_p(N^{-1/2})\) for \(j \in \mathcal{A}^c\).

The condition \(O_p(N^{-1/2}) \le \gamma_N\) holds with probability approaching 1, as Assumption \ref{assump:penalty} implies \(\sqrt{N} \gamma_N \to \infty\), so \(\gamma_N\) decays slower than \(N^{-1/2}\). Therefore, the KKT conditions are satisfied with \(\hat{\theta}_{N,j} = 0\) for all \(j \in \mathcal{A}^c\) with probability tending to 1, establishing sparsity recovery.

For the asymptotic normality of the active set, note that \(\hat{\theta}_{N,\mathcal{A}}\) is consistent for \(\theta_{\mathcal{A}}^*\) and \(\gamma_N \to 0\). Thus, for sufficiently large \(N\), \(\min_{j \in \mathcal{A}} |\theta_j^*| > a \gamma_N\), and by consistency, \(|\hat{\theta}_{N,j}| > a \gamma_N\) for \(j \in \mathcal{A}\) with probability approaching 1. By the SCAD-type penalty property in Assumption \ref{assump:penalty}, this implies \(p'_{\gamma_N}(|\hat{\theta}_{N,j}|) = 0\) for \(j \in \mathcal{A}\).

The KKT conditions for \(\mathcal{A}\) then reduce to
\begin{equation}
    \frac{\partial L_N(\hat{\theta}_{N,\mathcal{A}}, \mathbf{0}_{\mathcal{A}^c})}{\partial \theta_j} = 0, \quad j \in \mathcal{A},
\end{equation}

which matches the score equations for the oracle maximum likelihood estimator assuming knowledge of \(\mathcal{A}\).

A first-order Taylor expansion around \(\theta_{\mathcal{A}}^*\) yields
\begin{equation}
    \mathbf{0} \approx \nabla_{\theta_{\mathcal{A}}} L_N(\theta^*) + \nabla^2_{\theta_{\mathcal{A}}} L_N(\theta^*) (\hat{\theta}_{N,\mathcal{A}} - \theta_{\mathcal{A}}^*).
\end{equation}

Rearranging and scaling by \(\sqrt{N}\) gives
\begin{equation}
    \sqrt{N}(\hat{\theta}_{N,\mathcal{A}} - \theta_{\mathcal{A}}^*) \approx -(\nabla^2_{\theta_{\mathcal{A}}} L_N(\theta^*))^{-1} \sqrt{N} \nabla_{\theta_{\mathcal{A}}} L_N(\theta^*).
\end{equation}

By the law of large numbers for ergodic processes, \(-\nabla^2_{\theta_{\mathcal{A}}} L_N(\theta^*) \xrightarrow{p} \mathbf{I}_{\mathcal{A},\mathcal{A}}(\theta^*)\). By the central limit theorem for martingales,
\begin{equation}
    \sqrt{N} \nabla_{\theta_{\mathcal{A}}} L_N(\theta^*) = \frac{1}{\sqrt{N}} \sum_{i=1}^N \nabla_{\theta_{\mathcal{A}}} \ell_i(\theta^*) \xrightarrow{d} \mathcal{N}(0, \mathbf{I}_{\mathcal{A},\mathcal{A}}(\theta^*)).
\end{equation}

Applying Slutsky's theorem, we obtain
\begin{equation}
    \sqrt{N}(\hat{\theta}_{N,\mathcal{A}} - \theta_{\mathcal{A}}^*) \xrightarrow{d} \mathcal{N}(0, \mathbf{I}_{\mathcal{A},\mathcal{A}}^{-1}(\theta^*)),
\end{equation}
establishing asymptotic normality and the oracle property.

\section{Analysis of Computational Complexity and Deployment Feasibility}
\label{appendix:complexity_deployment}

A valid concern regarding our M$^3$TR framework is its apparent complexity, given the integration of multiple large pre-trained models and sophisticated custom modules. In this section, we provide a transparent and detailed analysis of the computational costs and argue for the framework's practical feasibility and scalability in a real-world, large-scale micro-video recommendation environment.

Our framework's architecture is predicated on a crucial design principle: \textbf{decoupling the expensive, large-scale data processing from the low-latency, real-time prediction task}. This is achieved through a two-phase process: (1) Offline Memory Bank Construction and (2) Online Real-time Inference.

\subsection{Phase 1: Offline Memory Bank Construction}
This phase involves pre-processing every video in the historical corpus to build the comprehensive Memory Bank ($B$). This is a one-time, upfront cost per video. The key operations are:

\begin{itemize}
    \item \textbf{Multi-Modal Feature Extraction:} For each video, we extract visual (ViT), acoustic (AST), and textual (AnglE) embeddings.  Using a cluster of GPUs, a corpus of millions of videos can be processed in batches with high throughput. The amortized cost per video is constant.
    
    \item \textbf{Temporal Embedding Generation:} This involves training the Mamba-Hawkes Process (MHP) on user interaction sequences and then using it to generate the temporal embedding $\mathbf{X}_i^t$. MHP training is the most novel and intensive part of this offline stage, but once trained, inference to produce the embeddings is fast. This step is also fully parallelizable across all videos.
    
    \item \textbf{Unified Retrieval Vector Generation:} We use powerful but heavy models like BLIP-2 and CLAP to generate rich textual descriptions from visual and audio content, which are then encoded into a unified semantic vector $\mathbf{R}_i$. This is computationally expensive but, again, a one-time offline cost that can be distributed across multiple machines.
\end{itemize}

\noindent \textbf{Practicality:} This offline batch-processing paradigm is standard industry practice for large-scale systems at companies like Google/YouTube, Meta, and ByteDance. The computational budget for offline data preparation is significant but considers a necessary investment to enable high-quality online services. Our approach fits perfectly within this established operational model.

To provide a concrete cost estimate, processing the MicroLens-100k training set (approx. 15,779 videos) on our experimental setup of 16 NVIDIA A100 GPUs has the following approximate wall-clock times:
\begin{itemize}
    \item Multi-Modal and Unified Vector Extraction (the most intensive part) requires approximately 12 hours of computation across the cluster.
    \item MHP inference, after a one-time training phase of $\sim$8 hours, is significantly faster, taking less than 2 hours for the entire set.
\end{itemize}

The total amortized cost is manageable and aligns with typical data-pre-processing pipelines in industry.
\subsection{Phase 2: Online Real-time Inference}
This is the latency-critical phase, which occurs when a new target video arrives for popularity prediction. The process is designed to be extremely lightweight:

\begin{itemize}
    \item \textbf{Feature Extraction (for Target Video):} The same feature extraction process from Phase 1 is applied to the single incoming video. On modern GPU hardware, processing a short micro-video through models like ViT and AST takes on the order of milliseconds.
    
    \item \textbf{Temporal-Aware Retrieval:} This step is the core of our system's efficiency. A naive search through millions of vectors in the Memory Bank would be prohibitively slow ($O(N)$, where $N$ is the bank size). Instead, we leverage state-of-the-art \textbf{Approximate Nearest Neighbor (ANN)} search algorithms and libraries, such as FAISS \citep{faiss} or ScaNN \citep{scann}. These systems use techniques like quantization and graph-based indexing to perform the retrieval in sub-linear, typically logarithmic, time ($O(\log N)$). For a bank of millions or even billions of videos, an ANN lookup takes only a few milliseconds.
    
    \item \textbf{Retrieval-Enhanced Fusion and Prediction:} After retrieving the top-S exemplars (where S is a small, fixed hyperparameter, e.g., $S=10$), the final prediction is made. The computational cost of this step is dominated by the cross-attention mechanisms between the target video's features and the S retrieved feature sets. The complexity is roughly $O(S \cdot d^2)$, where $d$ is the embedding dimension. Since both $S$ and $d$ are small constants, this cost is low, fixed, and, critically, \textit{independent of the total size of the Memory Bank}.
\end{itemize}

\subsection{Addressing Pipeline Brittleness} 
We acknowledge that maintaining a complex pipeline requires robust engineering practices. In a production environment, our framework would be implemented with the following considerations to ensure stability and robustness:
\begin{itemize}
    \item Model Versioning: Each component (e.g., ViT or BLIP-2) would be versioned. Updates to any single model would trigger the creation of a new, versioned Memory Bank, allowing for safe A/B testing without disrupting the live system.
    \item Monitoring and Alerting: Each stage of the pipeline would be monitored for data drift, concept drift, and performance degradation. For instance, we would track the distribution of output embeddings and the ANN retrieval latency.
    \item Fallback Strategies: In case of a failure in a complex component (e.g., the MHP module), the system can be designed to gracefully degrade. For example, it could fall back to a simpler retrieval mechanism that relies only on content similarity (SS\_modal), ensuring service availability.
\end{itemize}
\begin{table*}[h]
\centering
\resizebox{\textwidth}{!}{%
\begin{tabular}{@{}llll@{}}
\toprule
\textbf{Component} & \textbf{Computational Cost} & \textbf{Phase} & \textbf{Notes on Scalability} \\ \midrule
\multicolumn{4}{l}{\textbf{Memory Bank Construction}} \\
Multi-Modal Extraction (ViT, AST) & High (per video) & Offline & Embarrassingly parallel; scales with GPU cluster size. \\
MHP Temporal Embedding & Moderate (per video) & Offline & Fully parallelizable; amortized cost is low. \\
Unified Retrieval Vector (BLIP-2, CLAP) & Very High (per video) & Offline & One-time cost; standard practice for deep content understanding. \\
\midrule
\multicolumn{4}{l}{\textbf{Real-time Inference}} \\
Feature Extraction (Target Video) & Low & Online & Milliseconds on a single GPU. \\
ANN Retrieval (e.g., FAISS) & $O(\log N)$ & Online & Highly scalable; sub-linear time complexity w.r.t. Memory Bank size $N$. \\
Fusion \& Prediction (Cross-Attention) & $O(S \cdot d^2)$ & Online & Low and constant; independent of $N$. Dominant factor is small $S$. \\ \bottomrule
\end{tabular}%
}
\caption{Computational Analysis of M$^3$TR Components.}
\label{tab:complexity}
\end{table*}

\subsection{Our Framework's Superiority and Justification for Complexity}
The analysis in Table~\ref{tab:complexity} highlights the superiority of our design. We strategically accept a higher offline computational budget to achieve two critical goals for a real-world system:
\begin{itemize}
    \item \textbf{State-of-the-Art Accuracy:} As our experimental results demonstrate (e.g., up to 19.3\% nMSE improvement), the rich features generated by the ``heavy" offline models and the novel MHP module provide a significant lift in prediction accuracy over simpler methods. This level of performance is often a key business differentiator.
    \item \textbf{Low-Latency Online Performance:} By pre-computing all expensive representations and leveraging efficient ANN indexing, our online inference latency remains low and constant, making it fully suitable for user-facing applications where predictions are needed in real-time.
\end{itemize}
itemize
In conclusion, the complexity of M$^3$TR is not a liability but a deliberate and justified trade-off. It mirrors the engineering reality of modern large-scale recommendation systems, where offline processing power is leveraged to its fullest to provide fast, intelligent, and accurate online services. The framework is not only deployable but is architected in a way that is highly scalable and practical for the very environments it is designed to enhance.
\section{Baseline Models}
\label{models}
\begin{itemize}
\item \textbf{SVR}: Uses Gaussian kernel-based Support Vector Regression (SVR) to predict micro-video popularity.
\item \textbf{HyFea}: Employs the CatBoost tree model, leveraging multiple features (image, category, space-time, user profile, tags) for accurate prediction.

\item \textbf{Contextual-LSTM}: Integrates contextual features into the LSTM model, capturing long-range context to improve prediction accuracy.

\item \textbf{TMALL}: Introduces a common space to handle modality relatedness and limitations, enhancing popularity prediction.

\item \textbf{MASSL}: Utilizes a multi-modal variational auto-encoder model, capturing cross-modal correlations to predict popularity.

\item \textbf{MTFM}: Combines fuzzy trend matching and Informer in a multi-step prediction model, forecasting popularity trends.

\item \textbf{HMMVED}: Extracts and fuses multi-modal features through a variational information bottleneck for better prediction.

\item \textbf{CBAN}: Applies cross-modal bipolar attention mechanisms, effectively capturing correlations in multi-modal data to enhance prediction.

\item \textbf{MMRA}: Enhances prediction by retrieving relevant instances from a multi-modal memory bank and augmenting the prediction process.
\end{itemize}

\section{Datasets Details}
\label{Dataset}
MicroLens-100~\cite{ML100} is a large-scale micro-video recommendation dataset designed to address the challenge of the lack of publicly available large-scale datasets in the field of micro-video recommendation. The dataset contains approximately one billion user-item interaction records, 34 million users, and one million micro-videos. In addition to user interactions with micro-videos, MicroLens-100k includes various raw modalities of video information, such as titles, cover images, audio, and full-length videos.

As a content-driven benchmark for micro-video recommendation, MicroLens-100k enables researchers to leverage multiple modalities of video information for recommendation, rather than relying solely on item IDs or pre-trained network-extracted generic video features. This provides a reliable testing platform for the development of micro-video recommendation systems and fosters innovation and progress in the recommendation system field.

Besides, another dataset we used is provided by 2024 INFORMS Data Challenge committee. 
The existing datasets are listed in Table~\ref{dataset-comparison}, most of which are outdated and fail to reflect current multimodal, high-frequency user interactions, thus unable to realistically simulate real-world scenarios or accurately test algorithm performance. 

\begin{table}[h!]
    \centering
    \resizebox{\columnwidth}{!}{
    \begin{tabular}{cccccccc}
        \toprule
        \multirow{2}{*}{Dataset} & \multicolumn{5}{c}{Modality} & \multicolumn{2}{c}{Scale} \\
        \cmidrule(lr){2-6} \cmidrule(lr){7-8}
        & Text & r-Image & Audio & Video & User-feedback&\#User & \#Item \\
        \midrule
        Tenrec & \textcolor{red}{\faIcon{times}} & \textcolor{red}{\faIcon{times}} & \textcolor{red}{\faIcon{times}} & \textcolor{red}{\faIcon{times}} & \textcolor{red}{\faIcon{times}} & 6.41M & 4.11M \\
        Flickr & \textcolor{red}{\faIcon{times}} & \textcolor{red}{\faIcon{times}} & \textcolor{red}{\faIcon{times}} & \textcolor{red}{\faIcon{times}} & \textcolor{red}{\faIcon{times}} & 8K & 105K \\
        Pinterest & \textcolor{red}{\faIcon{times}} & \textcolor{green}{\faIcon{check}} & \textcolor{red}{\faIcon{times}} & \textcolor{red}{\faIcon{times}} & \textcolor{red}{\faIcon{times}} & 46K & 880K \\
        WikiMedia & \textcolor{red}{\faIcon{times}} & \textcolor{green}{\faIcon{check}} & \textcolor{red}{\faIcon{times}} & \textcolor{red}{\faIcon{times}} & \textcolor{red}{\faIcon{times}} & 1K & 10K \\
        Behance & \textcolor{red}{\faIcon{times}} & \textcolor{red}{\faIcon{times}} & \textcolor{red}{\faIcon{times}} & \textcolor{red}{\faIcon{times}} & \textcolor{red}{\faIcon{times}} & 63K & 179K \\
        KuaiRand & \textcolor{red}{\faIcon{times}} & \textcolor{red}{\faIcon{times}} & \textcolor{red}{\faIcon{times}} & \textcolor{red}{\faIcon{times}} & \textcolor{red}{\faIcon{times}} & 27K & 32.03M \\
        KuaiRec & \textcolor{red}{\faIcon{times}} & \textcolor{red}{\faIcon{times}} & \textcolor{red}{\faIcon{times}} & \textcolor{red}{\faIcon{times}} & \textcolor{red}{\faIcon{times}} & 7K & 11K \\
        Reasoner & \textcolor{green}{\faIcon{check}} & \textcolor{green}{\faIcon{check}} & \textcolor{red}{\faIcon{times}} & \textcolor{red}{\faIcon{times}} & \textcolor{red}{\faIcon{times}} & 3K & 5K \\
        \textbf{MicroLens} & \textcolor{green}{\faIcon{check}} & \textcolor{green}{\faIcon{check}} & \textcolor{green}{\faIcon{check}} & \textcolor{green}{\faIcon{check}} & \textcolor{green}{\faIcon{check}} & 30M & 1M \\        \textbf{INFORMS} & \textcolor{green}{\faIcon{check}} & \textcolor{green}{\faIcon{check}} & \textcolor{green}{\faIcon{check}} & \textcolor{green}{\faIcon{check}} & \textcolor{green}{\faIcon{check}} & $\sim$ 350 & 2.2K \\TikTokActions  & \textcolor{green}{\faIcon{check}} & \textcolor{red}{\faIcon{times}} & \textcolor{green}{\faIcon{check}} & \textcolor{green}{\faIcon{check}} & \textcolor{red}{\faIcon{times}}& & 283.5K \\
        Kaggle TikTok Video & \textcolor{green}{\faIcon{check}} & \textcolor{red}{\faIcon{times}} & \textcolor{red}{\faIcon{times}} & \textcolor{green}{\faIcon{check}} & \textcolor{green}{\faIcon{check}} & & \\
        Vine Dataset & \textcolor{green}{\faIcon{check}} & \textcolor{red}{\faIcon{times}} & \textcolor{green}{\faIcon{check}} & \textcolor{green}{\faIcon{check}} & \textcolor{green}{\faIcon{check}} & 98.2K & 303.2K \\
        News on TikTok & \textcolor{green}{\faIcon{check}} & \textcolor{red}{\faIcon{times}} & \textcolor{green}{\faIcon{check}} & \textcolor{green}{\faIcon{check}} & \textcolor{red}{\faIcon{times}} & & $>$4K \\
        Kaggle TikTokDataset & \textcolor{red}{\faIcon{times}} & \textcolor{green}{\faIcon{check}} & \textcolor{red}{\faIcon{times}} & \textcolor{green}{\faIcon{check}} & \textcolor{red}{\faIcon{times}} &  & $>$100K img \\
        YouTube Trending Videos & \textcolor{green}{\faIcon{check}} & \textcolor{red}{\faIcon{times}} & \textcolor{red}{\faIcon{times}} & \textcolor{green}{\faIcon{check}} & \textcolor{green}{\faIcon{check}} & & \\
        Snapchat Spotlight & \textcolor{green}{\faIcon{check}} & \textcolor{red}{\faIcon{times}} & \textcolor{green}{\faIcon{check}} & \textcolor{green}{\faIcon{check}} & \textcolor{green}{\faIcon{check}} & & 90K \\
        YouTube-8M & \textcolor{red}{\faIcon{times}} & \textcolor{red}{\faIcon{times}} & \textcolor{green}{\faIcon{check}} & \textcolor{green}{\faIcon{check}} & \textcolor{red}{\faIcon{times}} & & 6.1M \\
        Instagram Reels Dataset & \textcolor{green}{\faIcon{check}} & \textcolor{red}{\faIcon{times}} & \textcolor{red}{\faIcon{times}} & \textcolor{red}{\faIcon{times}} & \textcolor{red}{\faIcon{times}} & & \\
        HowTo100M & \textcolor{green}{\faIcon{check}} & \textcolor{red}{\faIcon{times}} & \textcolor{green}{\faIcon{check}} & \textcolor{green}{\faIcon{check}} & \textcolor{red}{\faIcon{times}} & & $>$1.2M \\
        WebVid & \textcolor{green}{\faIcon{check}} & \textcolor{red}{\faIcon{times}} & \textcolor{green}{\faIcon{check}} & \textcolor{green}{\faIcon{check}} & \textcolor{red}{\faIcon{times}} & & 2M-10M \\
        HD-VILA & \textcolor{green}{\faIcon{check}} & \textcolor{red}{\faIcon{times}} & \textcolor{green}{\faIcon{check}} & \textcolor{green}{\faIcon{check}} & \textcolor{red}{\faIcon{times}} & & 3.3M \\
        MMSum & \textcolor{green}{\faIcon{check}} & \textcolor{red}{\faIcon{times}} & \textcolor{green}{\faIcon{check}} & \textcolor{green}{\faIcon{check}} & \textcolor{red}{\faIcon{times}} & & 5.1K \\
        InternVid & \textcolor{green}{\faIcon{check}} & \textcolor{red}{\faIcon{times}} & \textcolor{green}{\faIcon{check}} & \textcolor{green}{\faIcon{check}} & \textcolor{red}{\faIcon{times}} & & 7M \\
        \bottomrule
    \end{tabular}
    }
    \caption{Dataset comparison. ``r-Image" refers to images with raw image pixels. ``Audio" and ``Video" mean the original full-length audio and video content.}
    \label{dataset-comparison}
\end{table}

\section{Evaluation Metrics Details}
\label{appendix:evaluation metrics}
The main evaluation metric $nMSE$ is defined as follows:
\begin{align}
    nMSE=\frac{1}{N\sigma_{y_i}^2}\sum_{i=1}^N(y_i-\hat{y}_i)^2,
\end{align}
where $N$ represent the total number of micro-video samples, $y_i$ and $\hat{y}_i$ are the target and predicted popularity score for the $i$th micro-video sample, and $\sigma_{y_i}$ is the standard deviation of the target popularity. \par
In addition, we measure Spearman's Rank Correlation (SRC) coefficient, Pearson linear correlation coefficient (PLCC), and Mean Absolute Error (MAE) as complementary metrics. SRC, PLCC, and MAE are defined as follows:
\begin{align}
    SRC=1-\frac{6\sum d_i^2}{N(N^2-1)},
\end{align}
\begin{align}
    MAE=\frac{1}{N}\sum_{i=1}^N\lvert{y_i-\hat{y}_i}\rvert,
\end{align}
\begin{align}
    PLCC=\frac1N\sum_{i=1}^N\frac1T\sum_{j=1}^T(\frac{y_i^j-\bar{y}_i}{\sigma_{y_i}})(\frac{\hat{y}_i^j-\bar{\hat y}_i}{\sigma_{\hat{y}_i}}),
\end{align}
where $d_i$ is the rank difference of a micro-video between the prediction and the popularity target. $\sigma_{y_i}$ and $\sigma_{\hat{y}_i}$ stand for the standard deviation of the target and predicted popularity sequences for the $i$-th micro-video sample.\par
 A higher SRC value indicates a stronger monotonic correlation between targets and predictions, a higher PLCC value indicates a higher model performance, while a lower nMSE or MAE indicates a more precise prediction of the model. 

\section{Implementation Details}
\label{implementation details}
\noindent\textbf{Device.}
All our experiments are conducted on a Linux server with 16 NVIDIA A100 Tensor Core GPUs using a multi-thread dataloader and 503G memory. \\
\noindent\textbf{Hyper-Parameter Settings.}
We adopt the Bayesian Optimization ~\cite{Bayes}, searching through the parameter hyper-space, and find the lowest nMSE when setting [learning\_rate, batch\_size, weight\_decay, dropout, $\alpha$, K]=[1e-5, 64, 0.001, 0, 0.8, 10]. We apply the above set of parameters to acquire our result and compare M$^3$TR with other baselines. Hyper-parameter optimization and sensitivity analysis have been discussed in detail in the main paper.

\paragraph{Hyper-Parameters Analysis}
After initial adjustments, the model's performance shows low sensitivity to $weight\_decay$, $batch\_size$ and $dropout$. So we focus on optimizing and analyzing other key parameters: the number of frames captured ($K$) and the balance between positive and negative attention ($\alpha$).\par

\begin{figure}[ht]
\centering
\includegraphics[width=1\columnwidth]{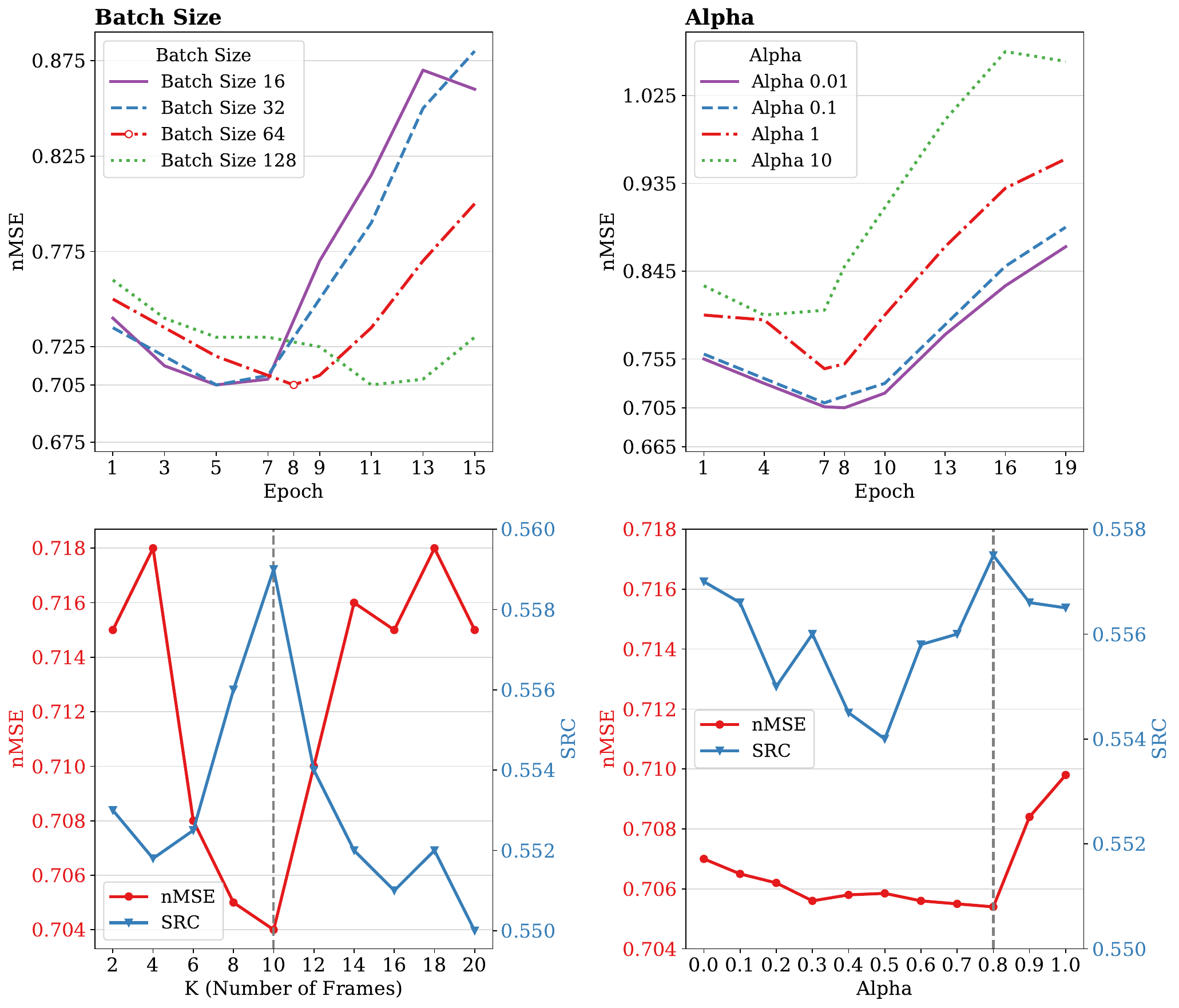}
\caption{Discussion with different hyper-parameters. The upper two figures show the nMSE during training with different batch sizes and $\alpha$s. The lower two figure illustrate the nMSE and SRC on testing dataset with different Ks and $\alpha$s.}
\Description{
A two-by-two set of hyper-parameter sensitivity plots. The upper-left
panel shows training nMSE across epochs for batch sizes 16, 32, 64,
and 128. The upper-right panel shows training nMSE across epochs for
four different alpha settings. The lower-left panel plots test nMSE
and SRC against the number of sampled video frames, with the two
metrics shown on separate vertical scales and a vertical reference
line marking the selected value K equals 10. The lower-right panel
plots the same two test metrics against alpha from 0 to 1, with a
reference line marking the selected value alpha equals 0.8. The lower
plots show modest metric fluctuations rather than a monotonic change,
illustrating relatively low sensitivity around the selected settings.
}
\label{fig7inter}
\end{figure}

% More videos aggregated in the retrieval process enable the model to pay more attention to the shared characteristics from similar videos and attain effective information, which ultimately improves the performance of M$^3$TR.
Figure \ref{fig7inter} illustrates the best nMSE and SRC obtained when training with different values of $K$ and $\alpha$. The model's performance is relatively insensitive to these two parameters. M$^3$TR performs best with $K = 10$ and $\alpha = 0.8$. The optimal values for $S$ and $lr$ are 10 and 1e-4, respectively. Overall, our model demonstrates strong robustness.

\section{More About the Mamba-Hawkes Process}
\label{sec:appendix_qualitative_mhp}
\subsection{Case Study: Disentangling the Mamba-Hawkes Process}
While quantitative results in the main paper demonstrate the overall superiority of our framework, a deeper qualitative analysis is essential to understand \textit{how} our Mamba-Hawkes Process (MHP) module operates and captures complex temporal dynamics. This case study aims to disentangle the contributions of the classical Hawkes component and the non-linear, Mamba-driven context vector, particularly in a challenging, non-monotonic popularity scenario.

We present a constructed yet representative case of a video that experiences an initial surge in positive engagement, followed by a sudden influx of controversial comments that subsequently suppresses further ``Like" activity. Such scenarios, where the valence of user interaction shifts, are notoriously difficult for conventional time-series models. Figure~\ref{fig:mhp_case_study} illustrates how our MHP module successfully navigates this complexity.

The visualization reveals the distinct roles of the MHP's components. The Classic Hawkes Intensity (green, dashed line) correctly models the self-exciting nature of Likes (blue markers) and Shares (green markers) during the initial phase (t=0 to t=9). Each positive event triggers a corresponding spike in the intensity, followed by an exponential decay. However, the fundamental limitation of this classical component is exposed in the second phase (t$>$9). It is oblivious to the semantic shift introduced by the wave of Comments (red markers) and, therefore, fails to predict the subsequent stagnation in user engagement.

This is where Mamba's Contribution becomes critical. By processing the entire event sequence through its selective state-space mechanism, the Mamba component learns the complex, cross-type event dependencies. Upon observing the influx of comments, it correctly infers a negative context, causing its output, $f_k(h(t))$, to become strongly negative, as shown by the red shaded area. This represents a sophisticated learned dynamic: a burst of controversial comments inhibits future like engagement. The Final MHP Intensity (blue, solid line), which is the synthesis of these two components, is thus corrected. The negative contribution from Mamba effectively counteracts the naive optimism of the classic Hawkes process, pulling the final predicted intensity down to accurately reflect the suppressed user interest.

In summary, this case study validates our core architectural hypothesis. The performance gains of M$^3$TR are not merely a result of using temporal data, but stem from the MHP's sophisticated ability to model intricate, non-linear interactions between different event types—a capability demonstrably beyond the reach of traditional point process models alone.

\subsection{Qualitative Analysis: MHP Dynamics Across Scenarios}
\label{sec:qualitative_analysis}
To complement our quantitative results and provide a more intuitive understanding of our model's capabilities, this section presents a qualitative analysis that visually elucidates the superiority of our Mamba-Hawkes Process (MHP) module. We compare the predicted cumulative popularity curves from our M\textsuperscript{3}TR model against a baseline (a classic Hawkes model) on four distinct and challenging popularity archetypes identified in real-world data.

\begin{figure*}[htbp]
    \centering
    \includegraphics[width=0.8\textwidth]{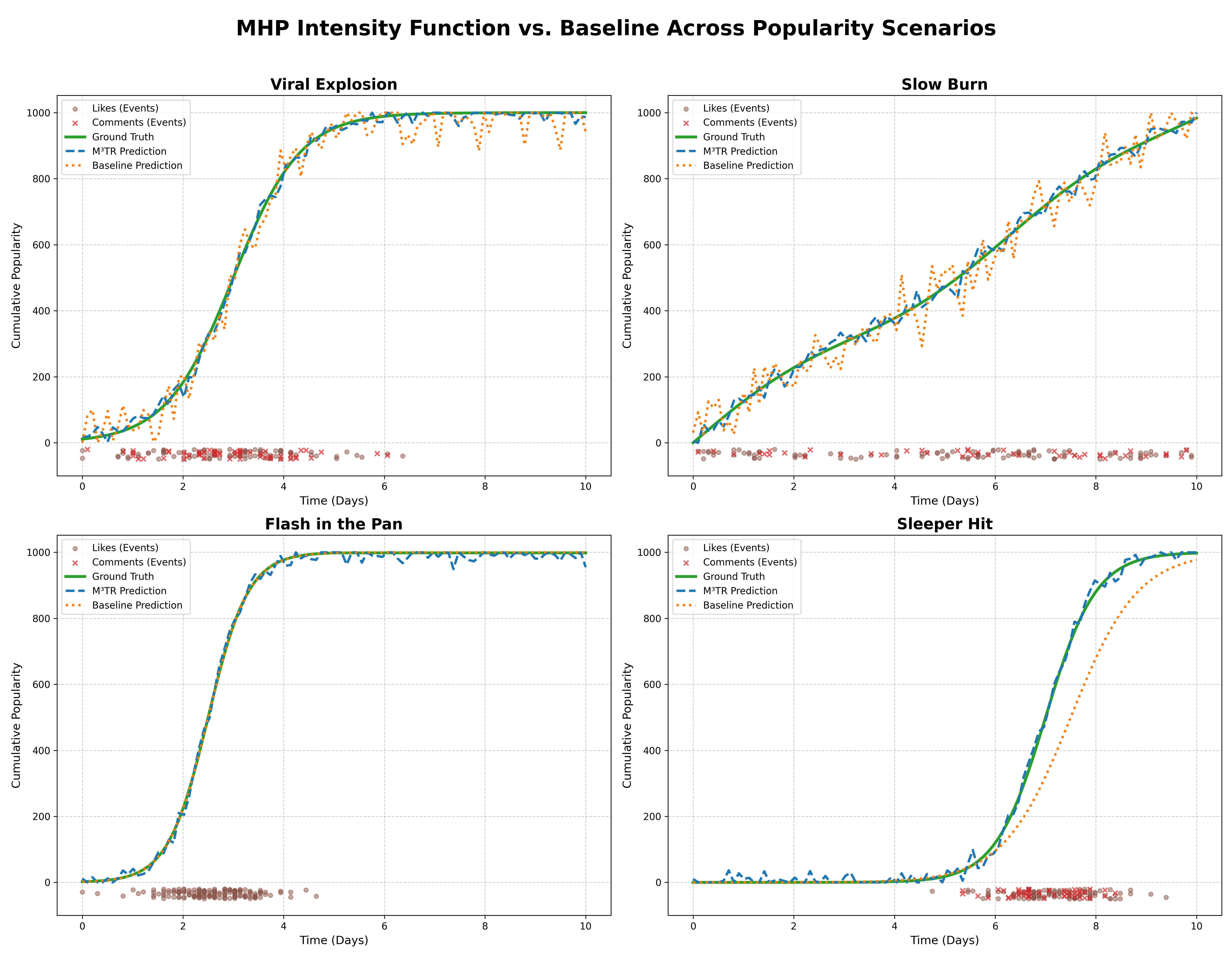}
    \caption{A comparative case study of cumulative popularity prediction across four scenarios. The plots show the Ground Truth (solid green), our M\textsuperscript{3}TR prediction (dashed blue), and a Baseline prediction (dotted orange). Scatter points at the bottom represent the timing of raw user interaction events (Likes and Comments). The ``Flash in the Pan" scenario particularly highlights M\textsuperscript{3}TR's ability to model trend reversals, a key limitation of the baseline.}
    \Description{Four line charts compare cumulative popularity over ten days for Viral Explosion, Slow Burn, Flash in the Pan, and Sleeper Hit scenarios. Each panel contains a ground-truth curve, an M3TR prediction, a baseline prediction, and event markers along the bottom for individual likes and comments. In Viral Explosion, popularity rises steeply between roughly days 2 and 5 and then saturates, with the M3TR curve closely following the ground truth while the baseline is visibly noisier. Slow Burn shows a gradual increase throughout the full period, again with M3TR closely tracking the underlying trend. Flash in the Pan shows a rapid rise followed by an early plateau, with M3TR nearly overlapping the ground-truth trajectory. Sleeper Hit remains near zero for about half of the observation period before a delayed rapid increase; M3TR follows the take-off closely, whereas the baseline rises later and lags behind the ground truth.}
    \label{fig:qualitative_scenarios}
\end{figure*}

Figure~\ref{fig:qualitative_scenarios} showcases this comparison. In relatively straightforward scenarios like ``Viral Explosion" and ``Slow Burn", both our M\textsuperscript{3}TR model and the baseline demonstrate competent tracking of the ground truth. Nevertheless, M\textsuperscript{3}TR's predictions maintain a consistently tighter fit, indicating higher precision and robustness.

The critical advantage of our MHP module is starkly revealed in the more complex, non-monotonic scenarios. For the ``Flash in the Pan" case, the baseline model, capable only of modeling self-excitation, incorrectly extrapolates the initial explosive growth, failing to recognize the subsequent stagnation in user engagement. In stark contrast, M\textsuperscript{3}TR accurately captures this abrupt turning point. This is because the MHP's Mamba component learns to interpret the influx of later-stage comments (the red ``x" markers) as a negative contextual shift, providing a crucial corrective signal to the overall intensity function. Similarly, in the ``Sleeper Hit" scenario, M\textsuperscript{3}TR exhibits superior responsiveness by correctly identifying the delayed take-off point and matching the steep growth trajectory, whereas the baseline model lags.

Collectively, these case studies visually corroborate our quantitative findings. They provide strong evidence that the sophisticated, context-aware architecture of our MHP module is essential for accurately forecasting the full spectrum of complex popularity trajectories, a task where simpler temporal models fundamentally fall short.

\end{document}